\documentclass[twocolumn,laps,pra,amsmath,amssymb]{revtex4}
\usepackage{amsmath}
\usepackage{graphicx}% Include figure files
\usepackage{epsfig}
\usepackage[usenames,dvipsnames]{color}
\usepackage{epstopdf} % comment this row
\usepackage{capt-of}

\definecolor{nblue}{rgb}{0.3,0.3,1.0}%229
\newcommand{\ea}{{\it et al.}}

\newcommand{\tmem}[1]{{\em #1\/}}
\newcommand{\tmop}[1]{\ensuremath{\operatorname{#1}}}

\begin{document}
\title{A case study in programming a quantum annealer for hard operational planning problems}
\author{Eleanor G. Rieffel, Davide Venturelli, Bryan O'Gorman, 
Minh B. Do, Elicia Prystay, and Vadim N. Smelyanskiy}
\affiliation{QuAIL, NASA Ames Research Center, Moffett Field, CA 94035}
% \author{Eleanor G. Rieffel${}^1$, Davide Venturelli${}^1$, Bryan O'Gorman${}^1$, \\
% Minh B. Do${}^1$, Elicia Prystay${}^1$, and Vadim N. Smelyanskiy${}^1$}
% \affiliation{QuAIL, NASA Ames Research Center, Moffett Field, CA 94035}
% \author{Eleanor G. Rieffel \and Davide Venturelli \and Bryan O'Gorman \and \\
% Minh Do \and Elicia Prystay \and Vadim Smelyanskiy\\
% NASA Ames Research Center\\
% Moffett Field, CA 94035\\
% }
% \maketitle

\begin{abstract}
We report on a case study in programming an early quantum annealer to
attack optimization problems related to operational planning. 
While a number of studies have looked at the performance of 
quantum annealers on problems native to their architecture, 
and others have examined performance of select problems stemming
from an application area, ours is one of the first studies of a
quantum annealer's performance on parametrized families of hard problems
from a practical domain. We explore
two different general mappings of planning problems to quadratic
unconstrained binary optimization (QUBO) problems, and apply them
to two parametrized families of planning problems, navigation-type 
and scheduling-type. We also examine two more compact, but problem-type
specific, mappings to QUBO, one for the navigation-type planning problems
and one for the scheduling-type planning problems. We study embedding
properties and parameter setting, and examine their effect on the efficiency 
with which the quantum annealer solves these problems. From these
results we derive insights useful for the programming and design of 
future quantum annealers: problem choice, the mapping used, the properties
of the embedding, and the annealing profile all matter, each significantly
affecting the performance. 
\end{abstract}

\maketitle

\section{Introduction}
\label{sec:intro}

Quantum computing has been proven to provide more efficient means 
for solving certain classes of specialized problems than is possible 
classically, and in other cases is strongly suspected to do so
\cite{RPbook,NCbook}. There are other classes of problems, however,
for which quantum computing does not provide an advantage. 
One of the biggest open questions in quantum computing is 
the breadth of its applications. 
% A significant impediment has been that
% quantum computational devices were not available. While researchers 
% have successfully performed theoretical analyses to prove advantages 
% over classical computation in some cases, researchers could not provide
% empirical evidence. 
Many of the most complex computations carried out
in the practical world today use heuristic algorithms which have not
been mathematically proven to outperform other approaches, but have 
been shown to be more effective empirically. Quantum heuristic algorithms
exist, but it is only when quantum computational devices that can carry 
out these algorithms become available that we can learn whether they 
are more effective than current classical approaches. The most prominent 
quantum heuristic is quantum annealing. 

Quantum annealing \cite{Farhi-98,Das08,Johnson11,Smelyanskiy12}  
is a metaheuristic for solving optimization problems
which bears some resemblance to simulated annealing, a classical 
metaheuristic. 
It works by starting the system in the ground state of a
known, easy-to-implement Hamiltonian $H_I$ and gradually varying
the Hamiltonian until it becomes a Hamiltonian $H_P$ that encodes the
cost function for the problem at hand:
\begin{equation}
H(s) = A(s)H_I + B(s) H_P,
\label{eqn:QAeqn}
\end{equation}
% where $s$ varies from $0$ to $1$
where $s\in [0,1]$, $A(0) = 1 = B(1)$, and $A(1) = 0 = B(0)$.
The intuition behind quantum annealing is that it explores the 
cost-function landscape, but has means of
exploration not open to classical methods such as quantum tunneling.

Within the last couple of years, D-Wave quantum annealers have become
available. While debate continues
as to the extent to which the D-Wave machine is quantum 
\cite{Johnson2011quantum,Boixo2013experimental,Smolin2013classical,Wang5837comment,Boixo2014evidence,Shin2014quantum,Vinci2014distinguishing,Shin2014comment},
these machines provide the first opportunity for researchers to 
experiment with quantum annealing. 
Quantum computational hardware is maturing to the point that a number
of different quantum computational devices implementing specialized
algorithms such as quantum annealing will become available in the
next several years. 

Because most physical interactions are $2$-local 
(i.e. pairwise), most emerging quantum technologies will support only
$2$-local interactions. For this reason, the problem Hamiltonian $H_P$
should be Ising, containing only $2$-local terms between the qubits. 
A standard translation maps between problem Hamiltonians on qubits
and cost functions of binary variables.
When translated to a cost function, the $2$-local condition on the
problem Hamiltionian  means that the cost function must be
a Quadratic Unconstrained Binary Optimization (QUBO) problem.
Furthermore, limitations on the ability to couple a single qubit to
more than a few other qubits, means that variables that appear in many 
terms will need to be represented by  multiple physical qubits in order
to enable the required connections. For example, the D-Wave processors
use a Chimera architecture in which each qubit is connected to at most 
$6$ other qubits (See Fig.~\ref{fig:ChimeraArch}), so any logical
qubit in problem Hamiltonian that appears with more than $6$ other
logical qubits must be represented by multiple physical qubits
in order for the problem to be expressible in this architecture.
The first step, obtaining the QUBO, is referred to as {\it mapping}
the problem to QUBO. The second step is referred to as {\it embedding}
the QUBO in the hardware. Problems that fit directly on the machine,
so do not require an embedding step, are referred to as {\it native} 
problems.

Given a specific quantum annealing hardware architecture,
there are three high-level research challenges:
\begin{itemize}
\item {\bf Problem design:} Identify potential applications with appropriately
difficult combinatorial optimization problems; extract core aspects of these
problems that contribute to their difficulty; generate families of
benchmark problems that are small enough to be run on the newly
available devices, but are nevertheless interesting in spite of 
their smallness. 
\item {\bf Mapping to QUBO: } Design general approaches to map these
problems to QUBO problems, and to make good choices of parameters, such 
as the relative weighting of QUBO terms.
\item {\bf Embedding in hardware: } Determine which physical 
qubits should represent each logical qubit, 
the strength of the internal couplings (between qubits 
representing the same logical qubit), and 
how to distribute the external couplings (between sets of qubits representing
coupled logical qubits). Thus, there are two aspects of embedding: the
topological aspect and the parameter-setting aspect.
\end{itemize}
In future architectures, there will be additional research challenges, such
as making good choices for the annealing time and the annealing profile
(the functions $A(s)$ and $B(s)$ determining the weighting of the driver
Hamiltonian and the problem Hamiltonian throughout the run). 

A previous paper \cite{Rieffel2014AAAI} focused on 
the design of two parametrized families of hard planning problems, 
navigation-type and scheduling-type, that capture aspects common to 
many real-world planning problems, exhibit exponential scaling in
hardness even at small sizes, and provide insights into 
state-of-the-art planning algorithms. These problems can be used to
investigate novel approaches to planning problems, as we do here
for quantum annealing. Since these problems are optimization versions
of NP-complete problems, we expect any approach, whether quantum or 
classical, to scale exponentially with problem size. In the classical
case, the slope of that exponential can be radically different from
algorithm to algorithm, and we expect the same to be true for
quantum heuristics. Some quantum heuristics will be better than others,
and there is the exciting possibility that quantum approaches could 
outperform classical heuristics on these or other hard combinatorial 
optimization problems. Such an algorithm would have significant
practical impact.

This paper focuses on the second and third programming challenges, 
mapping and embedding, for hard combinatorial optimization problems
that arise in operational planning. 
We explore multiple ways of mapping planning problems
to QUBO problems, explore embeddings of these problems, 
and provide comparisons of the effectiveness of the DWave Two machine housed
at NASA Ames in solving these problems under different mappings and
parameter choices. In particular, we examine two different approaches
for mapping general planning problems to QUBO problems, and apply both of
these mappings to the navigation-type family and the scheduling-type
family of planning problems of \cite{Rieffel2014AAAI}. 
In part because we are only able to embed
the smallest size problems in the D-Wave Two architecture under these 
mappings, we developed two more compact but problem-type specific 
embeddings, one
for the navigation-type problems and one for the scheduling-type 
problems. Ultimately, we are interested in the general approaches, since
only they would be applicable to real-world planning problems that
contain aspects of both navigation and scheduling that are not
easily separated. But at this early
stage we can learn from the behavior of the quantum annealer on
the problems obtained from the more specialized mappings.

We also explore embedding properties and parameter choices, 
and their relation to the efficiency with which 
the D-Wave Two solves these problems. For example, the internal
couplings between physical qubits representing the same logical
qubits must be set. Here, we look at the case in which all
internal couplings have the same strength $J_{int}$, and investigate 
how the performance is affected by variations in this value.
We are particularly interested in what these initial
results tell us about how different future machine architectures 
and different programming choices could affect the ability 
of these machines to solve these problems.

% \section{Related Work}
% \label{sec:relWork}

Our work is the first to explore the programming and performance of a
D-Wave machine on parametrized families of hard problems stemming from 
applications.
A number of studies have benchmarked D-Wave performance on native problems
\cite{Wang2013benchmarking,Wang2014probing,Boixo2014evidence,Wang2014performance,Ronnow2014defining}. 
A recent study \cite{Venturelli14} analyzes performance on 
families of non-native problems, random $2$-dimensional Ising lattices and
random Ising problems on fully connected graphs. As non-native problems,
they do require embeddings, and the paper explores similar issues in that 
context to some we explore here such as the setting of the internal
coupling constant $J_{int}$, but while these structures appear in
appications, these problems do not come from a specific applications. 
On the other hand, a number of groups have explored applications 
\cite{Kassal2010simulating,Perdomo2012finding,Babbush2012construction,Babbush2013adiabatic,PerdomoOrtiz14IEEE,OGorman14}, but either on set of
small instances for which there is no notion of hardness or a 
few larger, but specific
instances, rather than parametrized families of instances whose hardness
is expected to scale exponentially for all approaches, and on which the
performance of the best classical approaches confirms the exponentially
increasing hardness with size. Very recent work \cite{PerdomoOrtiz14QAProg} 
suggests a 
programmatic approach to some of the issues we discuss here, such as
setting $J_{int}$, with examples to problems from a few different 
application domains.
% \egr{Add that some of the application work used the blackbox approach
% and separate out those papers into two classes?}

Our main contributions include an analysis of performance 
on a large, parametrized set of hard benchmark problems stemming
from an application domain, a
comparison of the effectiveness of different mappings of these
problems to QUBO, an investigation 
of embedding properties and parameter setting and
their effect on performance, and an 
evaluation of future architectures in light of these findings.
% \begin{itemize}
% \item Evaluation on a large, parametrized set of benchmark problems.
% \item Comparison of the effectiveness of different mappings of these
% problems to QUBO.
% \item Analysis of embedding properties and parameter setting and
% their effect on performance. 
% \item Evaluation of future architectures in light of these findings.
% \item Insight into the programming and design of future quantum annealers.
% \end{itemize}
% We find, for example, that 
We derive from this study insights useful for the programming and design of
future quantum annealers. Specifically,
\begin{itemize}
\item scheduling-type planning problems are
more amenable to near-term quantum annealing approaches than
navigation-type planning problems, which are difficult to embed in 
the hardware, 
\item the choice of QUBO mapping makes a
marked difference in the success of an annealing algorithm even
when the QUBO sizes are similar, 
\item embedding metrics beyond embedding size and maximum component size 
are needed in order to predict and optimize performance, 
\item increasing the size of the unit cells of the Chimera architecture
(Fig.~\ref{fig:ChimeraArch}), and thereby the local
connectivity, would much more significantly impact our ability to 
run instances from applications than would simply increasing the number
of unit cells, and
\item support for different annealing profiles
(weightings of the driver and problem Hamiltonians throughout the run) 
need to be supported and could potentially lead to significantly 
improved results.
\end{itemize}

We first provide a brief review of classical planning 
(Sec.~\ref{sec:classical_planning}),
and then describe 
the parametrized families of hard navigation-type and scheduling-type 
planning problems used in our experiments in Secs.~\ref{sec:navigation} and
\ref{sec:scheduling} respectively. 
Readers who have previously read
\cite{Rieffel2014AAAI} may skip to section Sec.\ref{sec:qubos}.
In Sec.\ref{sec:qubos}, we describe the two general approaches 
to mapping classical planning problems to QUBO problems, followed 
by the two problem-type specific mappings.
Our methods, particularly the parameters we used for our annealing
runs, are described in Sec.\ref{sec:methods}.
Sec.\ref{sec:schResults} describes and analyzes our results on the
scheduling-type family of planning problems, and Sec.~\ref{sec:navResults}
describes our few results on the navigation-type family of planning problems.
Sec.\ref{sec:embResults} examines the embeddability of these problems
in future hardware architectures.
In Sec.\ref{sec:concl}, we conclude with a summary of our results and 
their implications for the programming and design of quantum annealers.

\section{An Overview of the Classical Planning Formalism}
\label{sec:classical_planning}

Classical planning problems, specifically STRIPS planning problems
\cite{Fikes72,Ghallab04}, are expressed in terms of binary 
{\it state variables} (sometimes called predicates) and {\it actions} 
(sometimes called operators). 
Examples of state variables in the rover domain are 
``Rover R is in location X" and ``Rover R has a soil sample from 
location X,'' which may be true or false. Actions consist of two 
lists, 
\begin{itemize}
\item a set of {\it preconditions} and 
\item a set of {\it effects} (or postconditions). 
\end{itemize}
The set of preconditions can be divided into positive preconditions,
those that must be true, and negative preconditions, those that must
be false. The set of effects can be divided similarly into a
set of positive effects and a set of negative effects.
In classical planning, it is conventional that the preconditions for
an action must be positive, so the set of preconditions is a subset of
state variables that must be set to true in order for the action to
be possible to carry out. When this convention is followed, the set
of negative preconditions will be empty. The effects of an action 
consists of a subset 
of state variables with the values they take on if the action is carried
out. For example, the action ``Rover R moves from location X to location Y"
has one precondition, ``Rover R is in location X = true" and has two
effects ``Rover R is in location X = false" and
``Rover R is in location Y = true."

A specific planning problem specifies an {\it initial state}, with values
specified for all state variables, and a {\it goal}, specified values
for one or more state variables. As for preconditions, goals are
conventionally positive, so the specified value for the goal variables 
is true. Generally, the goal specifies values
for only a small subset of the state variables. A plan is a sequence of
actions. A valid plan, or a solution to the planning problem, is
a sequence of actions $A_1, A_2, ..., A_L$ such that the state at 
time step $t_{i-1}$ meets the preconditions for action $A_i$, the
effects of action $A_i$ are reflected in the state at time
step $t_i$, and the
state at the end has all of the goal variables set to true.

\begin{figure*}[t]
\begin{center}
\includegraphics[width=0.9\textwidth]{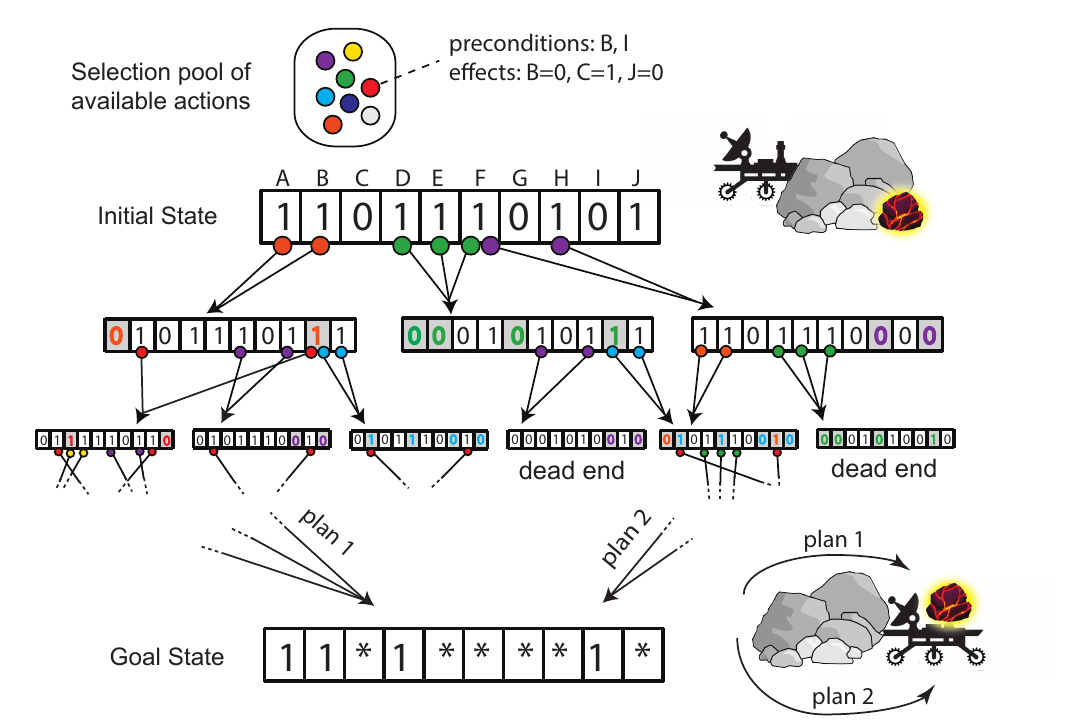}
\caption{
Pictorial view of a planning problem. The initial state (e.g. Rover to the left behind the rocks, without payload) is specified by assigning 
True (1) or False (0) to state variables (named A-J in this oversimplified example). The planning software navigates a tree, where a path represents a sequence (with possible repetitions) of actions selected from 
a pool (colors). Each action has preconditions on the state variables (e.g. moves can be done around the rocks and not through) which need to be satisfied in order for the actions to be executed (the circles
under the state variables in the search tree needs to match True=1) and has an effect on the state (colored variables in shaded regions of the new state have changed values). A valid search plan (multiple valid plans are possible) 
will reach the goal state (e.g. Rover in front of the rocks to the right, with a sample collected).
}
\label{fig:planningOverview}
\end{center}
\end{figure*}

\section{Parameterized Families of Navigation-Style Planning Problems}
\label{sec:navigation}

Navigation is a critical component in many planning 
applications and existing planning 
benchmarks \cite{Long03,helmert:2003,hoffmann:local-topology}.
Rover navigation is one such domain.
%the rover is dropped at a given location on a planet such as Mars. 
Given a list of locations a rover
must visit to, say, take picture or analyze samples,
the planner must find a route that makes optimal use of 
resources, such as time and power, satisfies multiple constraints, and 
achieves all goals. 
Under the assumptions that each location need be visited only once, 
the high-level navigation problem is similar to the Hamiltonian Path problems
we investigate. 
% Moreover, in many cases, the rover needs to go back to the lander to 
% transfer the data and collected samples, it can also resemble 
% Traveling Salesman or Hamilton Cycle problems.
% The parametrized families of navigation-type planning problems are based
% on directed Hamiltonian path (DHP) and undirected Hamiltonian path (UHP), 
% both are graph-theoretic NP-hard problems.
% (which we abbreviate as DHP and UHP, 
% respectively), the Hamiltonian cycle (HC) problems, the traveling 
% salesman problem (TSP).

%In the planning problems inspired by these Hamiltonian path problems, 
%% HC and TSP problems,
%there is an action associated with each node of the graph, the action
%of visiting that node. The preconditions and effects of an action 
%corresponding to a node are determined by edge structure
%between that node and the other nodes in the graph. An edge 
%originating from one node and ending in another corresponds to  
%allowing the action corresponding to the second node to follow the
%action corresponding to the first node.\\

\subsection{Planning problems from undirected Hamiltonian path (UHP)}
\label{sec:UHP}

The undirected Hamiltonian path (UHP) problem 
on an undirected graph $G(V,E)$, with $n$ vertices $V$ and a set of 
edges $E$, is to find a path that visits each node exactly once. 
A planning problem instance based on this graph
may be formulated as follows.
For each vertex $v$, there are:
\begin{itemize}
\item An action $a_v$ representing visiting $v$.
\item A `goal' state variable $s^g_v$ to indicate that $v$ needs to 
be visited; $s^g_v = T$ (true) means $v$ has been visited. 
\item An `internal' state variable $s^i_v$ represents whether or not $v$ has been visited.
Specifically, $s^i_v = T$ means $v$ has \emph{not} been visited while $s^i_v = F$ (false) means that
it has been visited. This variable ensures that each vertex can be visited at most once.
While including both $s^g_v$ and $s^i_v$ (which always
have opposite values) seems redundant, it is necessary because 
of the convention that allows only
positive action preconditions and goals.
\item An `external' state variable $s^e_v$ represents whether or not 
the vertex $v$ can currently be visited given the edge structure of the graph. 
Specifically, it is set to $T$ by an action $a_{v'}$ corresponding 
to visiting a vertex $v'$
that is connected to $v$ by an edge. Otherwise, it is set to $F$.
\end{itemize}

%The UHP domain contains $n$ actions, each with three associated variables: 
%a goal variable, an `internal' variable, and an `external' variable. Altogether 
%there will be $3n$ state variables consisting of $n$ goals, $n$ internal variables, and $n$ external variables.
%
%The action for a given site corresponds to the action of going to that
%site. The `internal' variable for this action 
%% corresponding to a site 
%can only be set by that action, and it is used to indicate that the site 
%has been visited. To ensure that this action can only take place after 
%an action corresponding to visiting one of its neighbors, 
%the `external' variable for this action is set by actions corresponding 
%to other sites that are connected by and edge to this site. 
%We now capture this intuitive explanation more formally in terms of
%preconditions and effects.

Each action $a_v$ has $2$ preconditions: (1) $s^i_v = T$, 
which indicates that this action has not been used in the plan already, 
and (2) $s^e_v = T$, indicating that this action can legally follow the previous action. 

Each action $a_v$ has $n+1$ effects: (1)  $s^g_v = T$, to indicate that $v$ has been visited,
(2) $s^i_v = F$, thus excluding $a_v$ 
from appearing twice in the plan,
%\footnote{ Notice that the internal variable always 
%has the opposite value to the goal variable for all sites. Classical STRIPS planning convention 
%requires that preconditions and goals must be positive. Thus, to enable both to be 
%positive, we must use two `opposite' variables.}. 
(3) sets each of the $n-1$ external variables $s^e_{v'}$ for each of the other
vertices $v'$: if there is an edge from $v$ to $v'$ then $s^e_{v'} = T$, enabling $a_{v'}$ to follow $a_v$;
if there is no edge from $v$ to $v'$ then $a_v$ sets $s^e_{v'} = F$, preventing $a_{v'}$ from following $a_v$.

%An observant reader will notice that the internal variable for a site
%always has the opposite value to the goal variable for the site, and
%may wonder why there is this duplication. Recall 
%classical planning convention that preconditions must
%be positive, as must the achievement of goals. To enable both to be
%positive, we must use two variables.

The initial state has all goal variables $s^g_v = F$ while all internal 
and external variables $s^i_v$ and $s^e_v$ have value $T$. Thus, any of 
the $n$ actions $a_v$ can be performed at the start.
A valid plan is a sequence of the $n$ actions that corresponds to 
a path along the edges that visits all vertices exactly once. \\
% Figure~\ref{fig:HamPathEx} shows one example.

\noindent{\bf Problem generation:} We obtain a parametrized family of 
UHP-based planning problems, parametrized by $n$ and $p$, by 
using the Erd\"os-R\'enyi model $G_{n,p}$ to randomly generate graphs
with $n$ vertices such that, for any pair of vertices, the 
edge between them is included with probability $p$.
We then derive planning problems from these graphs as described 
in the preceding paragraphs. We use the scaling parameter
$p=\left( \log n + \log \log n \right)/n$ that has been established
as at the phase transition for the closely
related Hamiltonian cycle problem \cite{Komlos83,Cheeseman91}.
We wrote a simple C++ program to generate these problems.

%To obtain a specific domain, we need to specify which external variables 
%each action sets to $1$ and which to $0$. For a domain with parameters 
%$n$ and $p$, each action $A_i$ sets the external variable for $A_j$ to $1$
%with probability $p$, and to $0$ with probability $1-p$. 
%For all $p$, there are $2^{n(n-1)}$ different domains, but unless $p = 1/2$,
%some are more likely than others.

\section{Parametrized families of scheduling-type planning problems}
\label{sec:scheduling}

% \noindent\emph{Graph-coloring \& Scheduling:} 
Many planning applications include scheduling 
aspects \cite{chien:spaceop2012}.
Scheduling, which deals with assigning resources and time to tasks 
while taking into account constraints, is in itself an important problem. 
Certain classes of scheduling problems correspond to graph coloring. 
% a well-studied combinatorial optimization problem. 
For example, a scheduling problem, with a set of tasks and
constraints that any pair of tasks competing for the same 
resource cannot be assigned the same time-slot, can be phrased as
a \emph{vertex coloring}, a well-known NP-complete problem. Specifically,
the chromatic number (the smallest number of colors needed) 
represents the smallest number of time-slots needed to complete
a corresponding schedule, thus representing the minimum makespan.

% \subsection{Planning problems from Graph Coloring (GC)}
\subsection{Planning problems from Vertex Coloring}

%A scheduling problem $P$ with a set of tasks $S$ in which there are 
%constraints that any pair of tasks $\{t_1, t_2 \}$ competing for the 
%same resource cannot be assigned the same time-slot can be mapped to 
%a vertex coloring problem for a graph $G(V,E)$ as follow:
%\begin{itemize}
%\item Each task $t \in S$ is represented as a vertex $v \in V$.
%\item Each time-slot is represented by a color.
%\item Each pair of tasks $\{t_1, t_2\}$ competing for a resource is represented by an edge $e = \{v_1, v_2 \}$ with $v_1, v_2$ representing $t_1, t_2$ accordingly.
%\end{itemize}
%The chromatic number of $G$ (smallest number of colors needed to color $G$) 
%represents the smallest number of time-slots needed for $P$, thus 
%representing the minimum makespan for $P$. 

Given an undirected graph $G = \{V, E \}$ with $n$ vertices,
% with a set of
%vertices $V$ and a set $E$ of edges generated as described by
%randomly assigning an edge between any two vertices with probability $p$, 
the planning problem to color $G$ with $k$ colors is formulated as follows. 
For each vertex $v$ there are:
\begin{itemize}
\item $k$ actions $a^c_v$ representing coloring $v$ with color $c$.
\item A `goal variable' $s^g_v$ representing whether or not $v$ has 
been colored at all.
\item A state variable $s^c_v$ representing whether or not $v$ has 
been colored with the color $c$.
\end{itemize}

%\footnote{In the PDDL representation, 
%the domain file uses $|V|$ action templates with color $c$ as parameters 
%and thus do not represent $|V| \times k$ (ground) actions directly.} 
%We restrict to the case in which 
%there are three colors, three time slots, available, so in our
%case there are $3n$ actions. Similarly, there are $4n$ state variables,
%four for each vertex $v$ corresponding to $colored(v)$, representing 
%whether or not $v$ is already colored, and $3$ variables $colored(v,c)$,
%representing whether or not $v$ is colored with color $c$. 

Let $C(v)$ be the set of neighboring vertices that are connected to $v$ by
an edge. For each action $a^c_v$, there are $|C(v)| + 1$ preconditions:
(1) $s^g_v = F$, which indicates that $v$ is not already colored;
and (2) for each $v_i \in C(v)$,  $s^c_{v_i} = F$, guaranteeing that 
none of neighboring $v_i$ are already colored with color $c$. 
% (preventing two connected nodes from being colored with the same color).

Each action $a^c_v$ has two effects: $s^g_v = T$ and $s^c_v= T$.

In the initial state, none of the vertices are colored: 
$\forall v \in V: s^g_v = F$, and $s^c_v= F$. 
The goal state requires that all vertices are colored: 
$\forall v \in V: s^g_v = T$. A plan is a sequence of $n$ actions, 
each of which colors a vertex $v$.\\

\noindent{\bf Problem generation:} As for the Hamiltonian path based problems, 
we obtain a parametrized family of graph-coloring-based planning problems
by randomly generating Erd\"os-R\'enyi graphs $G_{n,p}$
for a variety of values of $n$ and $p$.
A phase transition threshold in the $k$-colorability of $G(n,p)$
graphs has been established for all $k\geq 3$ in terms of the
parameter $c = m/n = p \times n$, the ratio of the number of edges to the
number of vertices \cite{Achlioptas99}.
The threshold scales as $c = k\log k$ in the leading
term, but the precise location of this threshold is still an open
question, even for $k = 3$ \cite{CojaOghlan13}.
Our runs were done with $c = 4.5$, a value intermediate to
the best current lower bound \cite{Achlioptas03}
and upper bound \cite{Dubois02} for the phase transition.
We extended Culberson et al.'s~\cite{Culberson95hidingour}
graph generator program, which provides methods to 
generate different types of graph controlled by various parameters,
to output PDDL files containing
the specification of planning problems derived from these graphs.

%EOF

\section{Mapping planning problems to QUBO form}
\label{sec:qubos}

The D-Wave quantum annealing machine can accept problems 
phrased in terms of a problem Hamiltonian $H_P$ in Ising form:
\begin{equation}
E_{\rm Ising}(s_1,\ldots,s_N)=-\sum_{i=1}^{N}h'_i s_i+\sum_{\langle i,j\rangle\in \,\texttt{E}}J'_{i,j} s_i s_j,\label{eq:Ising}
\end{equation} 
where $s_i=\pm 1$.
In traditional computer science, it is unusual to have variables $s_i$ 
whose values can be taken only from $\{-1, 1\}$, but it is common to 
have binary variables $z_i$ that take values from $\{0 , 1\}$. It is easy 
to convert between the two forms by taking $s_i=1-2z_i$. Any quadratic
function of variables $z_i$ can be converted to Ising form, up to a constant
which does not affect the energy minimization and so can be ignored: 
\begin{equation}
q(z_1,\ldots,z_N)=-\sum_{i=1}^{N}h_i z_i+\sum_{\langle i,j\rangle\in \,\texttt{E}}J_{i,j} z_i z_j,\label{eq:qubo}
\end{equation} 
Thus, it suffices to express any problem we want solved as a
Quadratic Unconstrained Binary Optimization (QUBO) problem
\cite{Choi08,Smelyanskiy12,Lucas13}, which will then be converted to 
Ising form to run on the D-Wave machine.

% \begin{center}
% \includegraphics[width=.7 \columnwidth]{./Figures/quantumAnnealingFigure.pdf}
% % \includegraphics[width=.45 \textwidth]{./Figures/quantumAnnealingFigure.pdf}
% \captionof{figure}{A schematic illustrating quantum annealing, including its
% capability to use quantum tunneling which is not available to classical
% approaches}
% \label{fig:QA}
% \end{center}

In this section, we describe two different mappings from general classical
planning problems, as described in Section \ref{sec:classical_planning},
to QUBO form. The first is a time-slice approach.
The second approach first maps a planning 
problem to a constraint satisfaction problem, and then reduces 
higher order terms to quadratic terms through a series of moves.
As we mentioned in Sec.~\ref{sec:classical_planning}, some classical 
planning algorithms follow the convention that an action can have only 
positive preconditions.
Since it is easy to do, we define our mappings generally, so that they
work with planning problems in which actions can have
negative preconditions, as well as those that follow the convention.
% Note that it would be wise to first simplify the planning problems
% before carrying out these mappings since quantum approaches would not 
% be bound by, for example, the constraints that the precondition and goal
% variables be positive.

\subsection{Time-slice method}

This mapping from general classical planning problems to QUBO form 
is a variant of the one developed and described in \cite{Smelyanskiy12}.

The method requires setting a specific plan length $L$. For the
two families of problems we consider, the plan length $L$ is
easy to determine; for the navigation-type problems, it is $n$, the
number of sites, and for scheduling-type problems, it is $1$, since
all vertices can be colored at the same time. In other cases, it
may be necessary to run quantum annealing on QUBOs corresponding to
different plan lengths, or to employ more sophisticated techniques
to determine the plan length to use \cite{blum:graphplan}.

\begin{figure}
%\begin{center}
\includegraphics[width=.9\columnwidth]{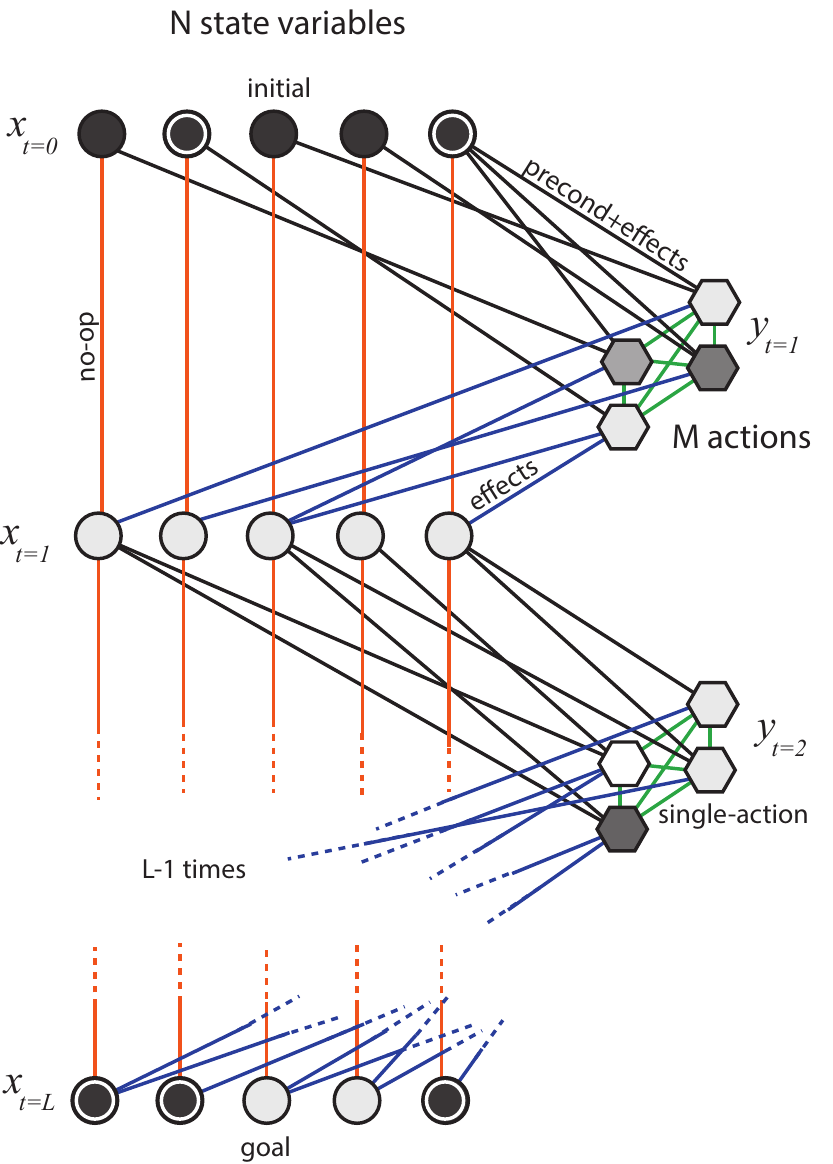}
\captionof{figure}{Time-slice QUBO structure for a 
planning problem with only positive preconditions and goals.
Each node represents a state 
variable (left) or an action (right) at any given time $t$. Time flows
from top to bottom, and variables $y_i^{(t)}$ for
the actions at time $t$ are shown between the
state variables $x_i^{(t-1)}$ for one time step and the
state variables $x_i^{(t)}$ for the next time step.
The node grayscale intensity represents 
the magnitude of local field (bias) $h_i$ applied to a given qubit $i$, 
and the double contour in a node indicates a negative bias.
% $h_i$ has a negative value. 
(One interesting property of this mapping is that the nodes 
representing state variables for $t\neq 0$ and $t\neq L$ 
have the same $h$ value, since they all come from $H_{no-op}$. For
this reason, they are all shown in same color in the diagram). Edges 
represents the couplings $J_{ij}$. Their weight is not 
illustrated in the figure. 
In this example we consider $H_{\tmop{single-action}}$ instead 
of $H_{\tmop{conflicts}}$, so all of the actions at a given time step are coupled to
each other.
}
\label{fig:qubo}
\end{figure}

If the original planning problem has $N$ state variables $x_i$ and
$M$ actions $y_j$ and we are looking
for a plan of length $L$, then we define a time-slice QUBO problem in terms 
of $N(L + 1) + LM$ binary variables. There are two groups of binary variables.
The first group consists of $N(L + 1)$ binary variables $x_i^{(t)}$ 
that indicate whether the state variable $x_i$ is $0$ or $1$ at time step $t$, 
for $t\in\{0,\dots,L\}$. % is the time index,
% and $i$ is the index of the state variable in the original planning problem.
% In addition, if the original planning problem has $M$ possible
% actions, we will have 
The second group consists of $LM$ binary variables $y_j^{(t)}$
that indicate whether or not the action $y_j$ is carried out between time 
steps $t-1$ and $t$.
% at time step $t$ or not. 
We can think of the entire set of binary variables as an alternating
string of $N$ variables corresponding to the state at a given time, followed
by $M$ variables corresponding to the actions, followed by a $N$ variables
corresponding to the state at the next time index, etc.
The structure of the QUBO is illustrated in Figure \ref{fig:qubo}.

% \end{center}

The total cost function is written as a sum
\begin{eqnarray*}
 H &=& H_{\tmop{initial}} + H_{\tmop{goal}} + H_{\tmop{no} - \tmop{op}}\\
 & & + H_{\tmop{precond}} + H_{\tmop{effects}} + H_{\tmop{conflicts}}. 
\end{eqnarray*}
We first give a mapping that is more general than we need, and then 
explain how it can be simplified in our situation. 
The first two terms are straightforward. They capture the initial
condition and the goal condition. 
Let $\mathcal{I}^{(+)}$ be the set of state variables that are 
$1$ in the initial condition and $\mathcal{I}^{(-)}$ be the set of 
state variables that are initially set to $0$. Similarly, let
$\mathcal{G}^{(+)}$ be the set of goal variables with value $1$ and 
$\mathcal{G}^{(-)}$ be the set of goal variables with value $0$.
(We describe the mapping for
general classical planning problems that do not necessarily follow
the convention that preconditions and goals must be positive.)
To capture the boundary conditions, the requirement that a plan start 
in the appropriate initial state and meets the goals, we include the following
terms in the cost function:
\[ H_{\tmop{initial}} = \sum_{i \in \mathcal{I}^{\left( + \right)}} \left( 1
   - x_i^{\left( 0 \right)} \right) + \sum_{i \in \mathcal{I}^{\left( -
   \right)}} x_i^{\left( 0 \right)} \]
and 
\[H_{\tmop{goal}} = \sum_{i \in \mathcal{G}^{\left( + \right)}} \left( 1 -
   x_i^{\left( L \right)} \right) + \sum_{i \in \mathcal{G}^{\left( -
   \right)}} x_i^{\left( L \right)}. \]

We next add terms to the cost function that penalize a plan if
an action is placed at time $t$ but the prior state does not have the
appropriate preconditions or if the subsequent state does not reflect
the effects of that action. Furthermore, we must penalize variable
changes that are not the result of an action. We start with this term,
the $H_{\tmop{no-op}}$ term, that penalizes variable changes:
\[ H_{\tmop{no} - \tmop{op}} = \sum_{t = 1}^L \sum_{i = 1}^N \left[
   x_i^{\left( t - 1 \right)} + x_i^{\left( t \right)} - 2 x_i^{\left( t - 1
   \right)} x_i^{\left( t \right)} \right] \]
This term gives cost penalty of 1 every time a variable is flipped.
Of course, when the effect of an action does result in a variable flipping,
we do not want this penalty, so we will make up for this penalty when
we add the term that corresponds to the effects of an action. First,
the term that penalizes violation of the preconditions looks like
\begin{eqnarray*}
H_{\tmop{precond}} &=& \sum_{t = 1}^L \sum_{j = 1}^M \left( \sum_{i \in
   \mathcal{C}_j^{\left( + \right)}} \left( 1 - x_i^{\left( t - 1 \right)}
   \right) y_j^{\left( t \right)} \right. \\
   & & + \left.\sum_{i \in C_j^{\left( - \right)}}
   x_i^{\left( t - 1 \right)} y_j^{\left( t \right)} \right) 
\end{eqnarray*}
where $\mathcal{C}_j^{(+)}$ is the set of positive preconditions for action
$j$ and $\mathcal{C}_j^{(-)}$ is the set of negative preconditions.
Let $\mathcal{E}_j^{(+)}$ be the set of positive effects for action
$j$ and $\mathcal{E}_j^{(-)}$ the set of negative effects.
The penalty if the appropriate effects do not follow the actions
is captured by the following term:
\begin{eqnarray*}
H_{\tmop{effects}}  &=&  \sum_{t = 1}^L \sum_{j = 1}^M \left( \sum_{i \in
		\mathcal{E}_j^{\left( + \right)}} y_j^{\left( t \right)} \left( 1 +
   x_i^{\left( t - 1 \right)} - 2 x_i^{\left( t \right)} \right) \right.\\
   & & + \left. \sum_{i \in
   \mathcal{E}_j^{\left( - \right)}} y_j^{\left( t \right)} \left( 2
   x_i^{\left( t \right)} - x_i^{\left( t - 1 \right)} \right) \right) .
\end{eqnarray*}
In order to understand this term, we must consider it together with
the no-op term. When $y_j^{\left( t \right)} = 1$, the corresponding 
term for $i \in \mathcal{E}_j^{\left( + \right)}$ 
(resp. $i \in \mathcal{E}_j^{\left( - \right)}$),
{\tmem{taken together with the no-op term}}, can be written 
$$\left( 1 + 2
x_i^{\left( t - 1 \right)} \right) \left( 1 - x_i^{\left( t \right)} \right)$$
(resp. 
$$\left( 3 - 2 x_i^{\left( t - 1 \right)} \right) x_i^{\left( t \right)}$$
for negative effects),
resulting in a positive penalty unless 
$x_i^{\left( t \right)} = 1$ (resp. $x_i^{\left( t \right)} = 0$). 
By using this form we have corrected for
the corresponding no-op term. 

Classical planners often allow for parallel plans not just
``linear plans'' in which more than one action can take place
at one time if those actions could have been done in any order, meaning that
the effects of any one action do not conflict with preconditions
of the other actions. 
What we have done so far works fine when the preconditions mean that
only one action can take place
per time period as is the case in the navigation problems. In the scheduling
problems, multiple actions can take place at the same time without 
conflicting. For general planning problems, we can either rule out 
multiple actions by imposing an additional term
$$H_{\tmop{single-action}} = 
\sum_{t = 1}^L \left(\sum_{j = 1}^M y_j^{(t)} - 1 \right)^2,$$
or we need to add terms to penalize potential conflicts.
To complicate matters, when more than one action can take place
at a given time, we are in danger of over-correcting for the no-op term.
If multiple actions at the same time have the same effect, the $H_{\tmop{effects}}$
term will add a term for each of those actions, thus overcompensating for the
no-op penalty. To avoid overcompensating, we penalize multiple actions at the
same time having the same effect, discouraging all such actions. A 
less stringent way to avoid overcompensating would be to add this 
penalty only when the effect
changes the variable, as we have done in the no-op term. The problem is
that natively that is not a quadratic term. Of course one could then
reduce that term, but here we choose to use the more stringent solution.
To ensure that two actions that conflict in the sense that
positive preconditions of one overlap with negative effects of the
other or vice versa, and to avoid overcompensating, we include the penalty
\begin{eqnarray*}
H_{\tmop{conflict}} &=& \sum_{t = 1}^L \sum_{i = 1}^N \left( \sum_{j|i \in
   \mathcal{C}_j^{\left( + \right)} \cup \mathcal{E}_j^{\left( - \right)}}
   \sum_{j' \neq j|i \in \mathcal{E}_{j'}^{\left( - \right)}} y_j^{\left( t
   \right)} y_{j'}^{\left( t \right)}\right. \\
   & & + \left.\sum_{j \left| i \in
   \mathcal{C}_j^{\left( - \right)} \cup \mathcal{E}_j^{\left( + \right)}
   \right.} \sum_{j' \neq j \left| i \in \mathcal{E}_{j'}^{\left( + \right)}
   \right.} y_j^{\left( t \right)} y_{j'}^{\left( t \right)} \right) . 
\end{eqnarray*}

While for explanatory purposes it was useful to include variables for the
state at time $t=0$, those can be set ahead of time, so that we don't
need to include the $H_{\tmop{initial}}$ term. The same is true
of the $H_{\tmop{goal}}$ term.
We can also replace all of their occurrences in 
$H_{\tmop{no-op}}$, $H_{\tmop{condition}}$, and $H_{\tmop{effect}}$
with these set values to simplify those constraints. 
Furthermore, if a state variable first appears at time step $t$,
then in the no-op term connecting it with the previous level, we
can set it to $0$. 
Finally, for the final action time slot, we can remove
any actions whose effects conflict with the goals or do
not contribute to the goals. 
These simplifications result in modified terms
$H'_{\tmop{no-op}}$, $H'_{\tmop{condition}}$, and $H'_{\tmop{effect}}$.
Additionally, since in our setting we have followed the convention that 
preconditions must be positive, we can use a simpler version 
of the $H_{\tmop{precond}}$ term:
% and goals 
%for the corresponding terms:
% \[H'_{\tmop{goal}} = \sum_{i \in \mathcal{G}^{\left( + \right)}} \left( 1 -
%   x_i^{\left( L \right)} \right) \]
\begin{eqnarray*}
H'_{\tmop{precond}} &=& \sum_{t = 1}^L \sum_{j = 1}^M \sum_{i \in
   \mathcal{C}_j^{\left( + \right)}} \left( 1 - x_i^{\left( t - 1 \right)}
   \right) y_j^{\left( t \right)}.
% H'_{\tmop{precond}} &=& \sum_{t = 1}^L \sum_{j = 1}^M \sum_{i \in
%   \mathcal{C}_j^{\left( + \right)}} \left( 1 - x_i^{\left( t - 1 \right)}
%   \right) y_j^{\left( t \right)}.
\end{eqnarray*}
For the navigation problems, the QUBO simplifies to
$$ H =  H'_{\tmop{no} - \tmop{op}}
  + H'_{\tmop{precond}} + H'_{\tmop{effects}},$$ 
% $$ H =  H'_{\tmop{goal}} + H_{\tmop{no} - \tmop{op}}
  %+ H'_{\tmop{precond}} + H_{\tmop{effects}},$$ 
% \begin{eqnarray*}
 % H &=&  H'_{\tmop{goal}} + H'_{\tmop{no} - \tmop{op}}\\
 % & & + H'_{\tmop{precond}} + H'_{\tmop{effects}}, 
% \end{eqnarray*}
and for the scheduling problems the QUBO simplifies to
\begin{eqnarray*}
 H &=& H_{\tmop{no} - \tmop{op}}\\
 & & + H'_{\tmop{precond}} + H_{\tmop{effects}} + H_{\tmop{single-action}}. 
% H &=& H'_{\tmop{goal}} + H_{\tmop{no} - \tmop{op}}\\
%  & & + H'_{\tmop{precond}} + H_{\tmop{effects}} + H_{\tmop{single-action}}. 
\end{eqnarray*}
% or, if we would like to allow multiple actions at the same time
% we can replace $H_{\tmop{single-action}}$ with $H_{\tmop{conflict}}$. 

\subsection{CNF approach} 

A CNF expression over $n$ Boolean variables $x_i$ 
consists of a bunch of clauses $C_a$ consisting of $k$ variables, possibly
negated, connected by logical {\sc or}s:
$$b_1 \vee b_2 \vee \dots \vee b_k$$
where 
$$b_i\in\{x_1, x_2, ..., x_n, \neg x_1, \neg x_2, ..., \neg x_n\},$$ 
and the number of variables $k$ in the clause can vary from clause to clause. 
A CNF for a $k$-SAT expression consists of clauses that all have 
the same number of variables $k$. In a CNF coming from $2$-SAT, 
for instance, all clauses have the form $b_1 \vee b_2$. 
In a CNF, all of the clauses must be satisfied,
which means they are connected by an {\sc and} operator (the reason for 
the ``conjunctive'' in ``conjunctive normal form"). 
% An example of a CNF expression
% consisting of $L$ clauses connected by logical {\sc and}s is
% \begin{eqnarray*}
% C_1 \wedge C_2 \wedge ...\wedge C_L &=& 
% (b^{(1)}_1 \vee b^{(1)}_2 \vee \dots \vee b^{(1)}_{k_1}) \wedge \\
% & & (b^{(2)}_1 \vee b^{(2)}_2 \vee \dots \vee b^{(2)}_{k_2}) \wedge \\
% & & \dots\wedge
% (b^{(L)}_1 \vee b^{(L)}_2 \vee \dots \vee b^{(L)}_{k_L}).
% \end{eqnarray*}

We used the first of the four PDDL to CNF translators built in to 
the SATPLAN planner \cite{kautz2004satplan04}, a classical SAT-based planner, 
to output planning problems in conjunctive normal
form (CNF). Compilation planners such as SATPLAN perform some preprocessing,
such as ``reachability'' and ``relevance'' analysis, 
as part of the translation to reduce the size of the output. 
Reachability analysis makes a quick determination as to whether a given
action can appear at a given time step, and removes from consideration at
that time step any actions that it has determined cannot be carried out,
which results in a simplified CNF expression. Similarly, if the possible 
actions at previous time steps cannot change a state variable, the
resulting expression is simplified accordingly. For example, in the first
time step, only actions whose preconditions are satisfied in the initial
state are considered. Then, only values of state variables that occurred
in the initial state or are effects of actions considered at the first 
time step are considered. This process is iterated until it no longer 
results in simpification. Relevance analysis is a similar process
that starts its analysis in the last time step, considering only state
variables that are goals, and then working backwards leaving only actions
with effects relevant to these goals in the last action time step. 
Relevant state variables at time step $t$ represent the union of 
state variables at time step $t+1$ and the preconditions of all 
relevant actions at time step $t$. SATPLAN's ``action-based'' encoding
uses an advanced reachability and relevance heuristic
analysis, and then further removes all variables representing state
variables while adding constraints that capture the relationships
between actions in consecutive time steps that were previously enforced 
by relationships between actions and state variables.

We convert a CNF instance to QUBO by first transforming it to 
Polynomial Unconstrained Binary Optimization (PUBO), a generalization 
of QUBO in which the objective function is a pseudo-Boolean 
of arbitrary degree. For each clause in a given CNF instance, 
we introduce a term to the PUBO instance equal to the conjunction 
of the negation of the literals in that clause, where a positive 
literal is replaced by the corresponding binary variable and a 
negative literal is replaced by the difference of one and the 
corresponding binary variable. For example, the CNF term 
$(x_1 \lor \lnot x_2 \lor \lnot x_3 \lor x_4)$ would correspond 
to the PUBO term $(1-x_1) x_2 x_3 (1-x_4)$.

We then reduce higher degree terms in the PUBO instance using an 
iterative greedy algorithm that is related to one described in
\cite{boros2002pseudo}. 
At each step, the pair of variables that appears 
in the most terms is replaced by an ancilla variable 
corresponding to their conjunction. (If there are multiple such pairs,
then one is chosen arbitrarily.) 
A penalty term is introduced 
to enforce that the ancilla variable indeed corresponds to the 
requisite conjunction. The penalty weight we use is equal to one plus 
the greater of the sums of the magnitudes of the positive coefficients 
and negative coefficients of the terms the ancilla is used to reduce 
\cite{babbush2013resource}.
The one is added to ensure that the constraint-satisfying states
have lower energy than the constraint-violating states. One is
convenient, and in keeping with the integer coefficients for the 
other terms, but any positive constant would do.
This procedure is repeated until the resulting PUBO is quadratic.

\subsection{Direct mapping of underlying graph problems to QUBO}

We now turn to two more compact, but problem-type specific mappings, the
first mapping navigation-type problems to QUBO problems and the second 
mapping scheduling-type problems to QUBO.

\subsubsection{Direct mapping of graph coloring to QUBO}

For a graph coloring problem with $n$ vertices and $k$ colors, 
we have $kn$ binary variables, $x_{ic}$, where $x_{ic} = 1$ means
that vertex $i$ is colored with color $c$, and $x_{ic} = 0$ means
it is not.

The QUBO contains two different types of penalty terms. The first 
corresponds to the constraint that each vertex must be colored
by one, and no more than one, color: $\sum_{c = 1}^k x_{ic} = 1$.
So for each vertex $i$, we have a term
$$\left(1 - \sum_{c = 1}^k x_{ic}\right)^2.$$
The second corresponds to the constraint that two vertices connected
by an edge cannot be colored with the same color. For each vertex $i$,
we have a term
$$\sum_{(i,j)\in E} \sum_c  x_{ic}x_{jc}.$$
Altogether the QUBO is
$$\sum_{i=1}^n\left(1 - \sum_{c = 1}^k x_{ic}\right)^2 +
\sum_{(i,j)\in E} \sum_c x_{ic}x_{jc}.$$

\subsubsection{Direct mapping of Hamiltonian Path Problems to QUBO}

For a Hamiltonian path problem with $n$ sites, we have $n^2$ variables
$$\{x_{11}, \dots, x_{1n}, x_{21}, \dots, x_{2n}, \dots, x_{n1}, \dots, x_{nn}.\}$$ 
The first index $i$ gives the site, the second index gives the time slot,
so $x_{ij} = 1$ means that the $i$th site is the $j$th site
visited, and $x_{ij} = 0$ means that the $i$th site is not visited in the
$j$th time slot.

There are three types of terms in the QUBO cost function: penalties if a site
is visited more or less than once, penalities if more than one site is visited
in a given time slot, and penalities for violation of edge constraints.

The first type of term enforceses  that each site is visited exactly once: 
For each site $i$, we will have a term of the form
$$(\sum_j x_{ij} - 1)^2.$$

The second type of term enforces that in each time slot no more 
than one site is visited 
(we may as well enforce it to be exactly one): For each time slot $j$,
we have a term of the form 
$$(\sum_i x_{ij} - 1)^2.$$

The third type of term is a single term
penalizing the violation of edge constraints. It penalizes
bisiting the $i'$th site right after the $i$th site if they
are not connected by an edge:
$$\sum_{j = 1}^{j = n-1} \sum_{i,i' st (i,i')\notin E} x_{ij} x_{i', j+1}.$$

There are $2n + 1$ terms all together.

\section{Methods}
\label{sec:methods}

All quantum annealing runs were performed on the $509$-qubit D-Wave Two 
machine 
housed at NASA Ames. In all cases, we used an annealing time of $20$ $\mu$sec, 
the fastest annealing time currently available on the machine, which is also
D-Wave's recommended annealing time given that current data suggests that an
even faster annealing time would be optimal \cite{Ronnow2014defining}. 
For each embedded QUBO 
instance, we performed $45,000$ anneals at each of ten gauges, 
for a total of $450,000$ anneals per QUBO instance. 
Gauges, which determine whether the bit values $\{0,1\}$ of each QUBO 
variable are mapped to $\{-1,1\}$ or $\{1,-1\}$, are used to reduce 
the effects of asymmetries in the hardware
\cite{Ronnow2014defining,PerdomoOrtiz14QAProg}.

From a QUBO instance generated as described in Section \ref{sec:qubos},
we generate the embedded instance by running D-Wave's heuristic embedding
software \cite{Cai-14} on the original QUBO instance. 
We use the software's default 
parameters, including a maximum of $10$ tries, unless otherwise noted.
The output of the embedding software is
a set of disjoint connected components of the hardware graph $C_i$, one
for each variable $x_i$ in the original QUBO.
We performed our own parameter setting, rather than using 
D-Wave's which tries successive parameter values, so that we could 
obtain statistics for a variety of parameter settings. 

From the original QUBO instance, we obtain the logical Ising instance 
$$h_i s_i + J_{ij} s_i s_j$$
through the standard translation of the variables $s_i=1-2z_i$.
To obtain the embedded Ising instance, we evenly distribute the bias
$h_i$ in the logical Ising instance among the qubits corresponding 
to the nodes in $C_i$: we set the 
linear coefficient for each variable $y$ in $C_i$ to be $\frac{h_i}{|C_i|}$.
We set all internal couplings, couplings between
physical qubits that represent the same logical variable in the
original QUBO, to the internal coupling constant $J_{int}$, a value we specify; 
the coefficient of all quadratic terms $yy'$ such that $y$ and $y'$ are 
both in $C_i$ for some $i$ are set to $J_{int}$.
We describe shortly the results we obtained in experiments 
varying this value. 
The only other couplings are between sets of physical qubits $C_i$ and $C_j$
representing two different logical variables $x_i$ and $x_j$ that appear 
together in a quadratic term in the original QUBO.
In many cases there is only one edge in the hardware graph between the qubits
in $C_i$ and $C_j$.
 %where $s_i$ and $s_j$ are coupled in the logical Ising.
In this case, we set the coupling between them to $J_{i,j}$. 
When more than one edge exists between these sets, we
choose one of them, and set its coupling to $J_{i,j}$. All other 
couplings are set to zero.

Our test set consists of $100$ solvable problem instances at the phase 
transition for each size for both planning problem types. We generated
the problems as described in \cite{Rieffel2014AAAI}, and then 
took the first $100$ solvable problems. 
For the sizes in which \cite{Rieffel2014AAAI} already had generated
problems, we use the first $100$ solvable instances tested there.
The smallest size problems we considered of the navigation-type were
of size $4$. For the scheduling-type problems, we started at size $8$ 
because the phase transition parameter is inaccurate for smaller sizes,
biasing the results toward unsolvable instances.

Because all of the problems we consider are solvable, we know the
ground state energy in all cases; zero, the minimal value of the QUBO
in all cases is attainable, and from that we can compute the ground state
energy of the embedded Ising problem that was actually run. For each 
embedded instance, once we obtain the $450,000$ results from the run, 
we check how many times the ground state energy was obtained, which
gives us the probability of solution $r$ for a $20$ $\mu$sec anneal. We
then compute the expected number of runs $k = \frac{\ln(1 - 0.99)}{\ln(1-r)}$ 
required to obtain a $99\%$ success
probability, multiply by the anneal time of $20$ $\mu$sec, and
report $20k$ $\mu$sec, the expected total anneal time 
to obtain a $99\%$ success probability. We are effectively using a 
$0.9$ sec.~cutoff time, since the expected anneal time when only one 
anneal solves is $0.9$ secs. Given that classical planners solve
these problems in less than $0.1$ secs., with the best planners
for these problems solving them in less than 
$0.01$ secs.~\cite{Rieffel2014AAAI}, this cutoff time seems reasonable.

We report the median 
expected total anneal time across $100$ instances, with error bars
corresponding to the $35$th and $65$th percentiles. Thus each data
point shown represents $45$ million anneals. While the total annealing
time for each point is only $90$ seconds, because the read-in and read-out
take considerably longer than the anneal time, and because of contention
for the machine, the wall clock time to obtain a single data point
is hours not minutes. Finding the embedding, by far the longest step
in the process, can take minutes for the largest instances, but fortunately
needs to be performed only once per QUBO instance.

The ground state energy of the embedded Ising model will not be obtained
if the final bit values differ between any two physical qubits representing
the same logical qubit. This observation suggests a simple, totally 
classical error correction scheme that uses majority voting among
all physical qubits representing the same logical qubit.
This error correction scheme is working at a different level from that
in Pudenz \ea\ \cite{Pudenz2013error}; the benchmark problems in 
Pudenz \ea\ are native problems that do not require an embedding step 
in order to be run on the hardware.
We report expected total anneal time both with and 
without this simple form of error correction.

\section{Results on scheduling-type planning problems}
\label{sec:schResults}

Because the scheduling-type planning problems embedded much more easily 
than the navigation-type planning problems, we were able to do significantly
more analysis of the the choices affecting the D-Wave Two's performance on 
these problems than on the navigation-type problems.
We first examine performance of the D-Wave Two on 
scheduling-type instances mapped using
the two general approaches for mapping planning
problems to QUBOs. We then turn to the results on these instances using
the mapping specific
to the scheduling approach, the direct mapping from scheduling-type
planning problems to QUBO. 

\subsection{Time-slice and CNF mapping results}

\begin{figure}
\includegraphics[width=\columnwidth]{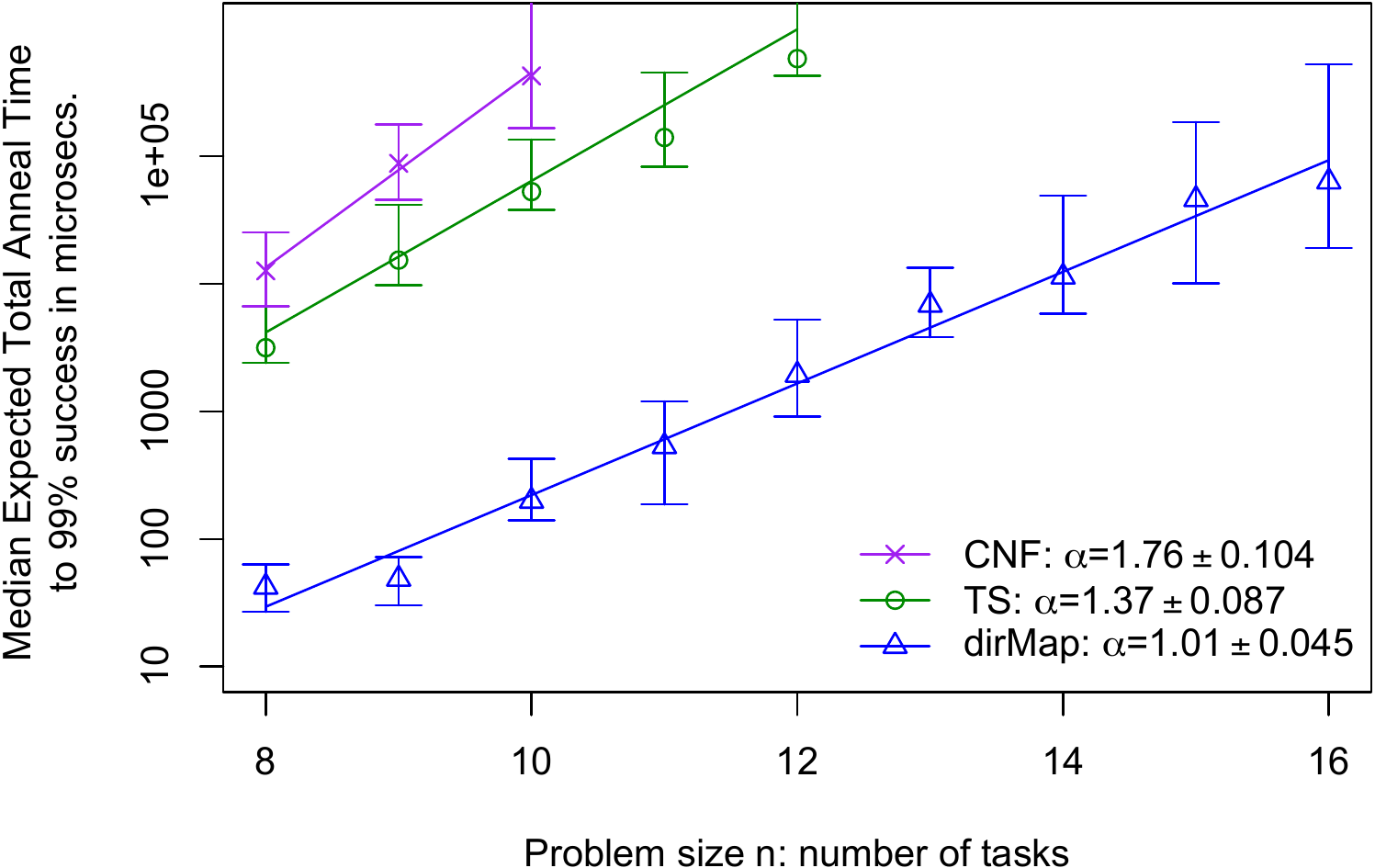}
\caption{{\bf Comparison of the median expected total anneal time to $99\%$ 
percent success for the three mappings of scheduling-type planning problems.} 
The results with the best $J_{int}$ are shown. 
(See Sec.~\ref{sec:Jint} and Fig.~\ref{fig:Jint} for how these
values were determined.)
For the CNF mapping, $J_{int}$ was $\{1.2, 1.3, 1.4\}$ for problem sizes
$\{8, 9, 10\}$.
For the time-slice mapping, $J_{int}$ was $\{1.6, 1.7, 1.7, 1.7, 1.6\}$ 
for problem sizes $\{8, 9, 10, 11, 12\}$. For the direct mapping,
$J_{int}$ was $\{1.1, 1.1, 1.3, 1.3, 1.3, 1.4, 1.4, 1.4, 1.4\}$ for problem
sizes $\{8, \dots, 16\}$. Each data point shows the median expected total annealing 
time to achieve $99\%$ success over the $100$ problems of each size given
on the x-axis. The error bars are at the 35th and 65th percentiles.
When at least half of the instances do not solve 
within the $0.9$ sec.~effective cutoff time, we no longer show the point. 
Also, when fewer than $65\%$ solve, the top of the error bar is 
indeterminate, as happened for the last point shown in both the CNF 
and time-slice series.
}
\label{fig:bestComp}
\end{figure}

Fig.~\ref{fig:bestComp} shows the relative performance, in terms
of median expected total annealing time for $99\%$ percent success, 
of the D-Wave Two on 
the benchmark set of scheduling-type planning problems. 
When at least half of the instances do not solve 
within the $0.9$ sec.~effective cutoff time, we no longer show the point. 
For the CNF mapping, that happens already by problem size $11$. 
For the time-slice instances,
at least half do not solve within the cutoff time by problem size $13$,
and for the direct map by problem size $17$.

\begin{figure}
\includegraphics[width=\columnwidth]{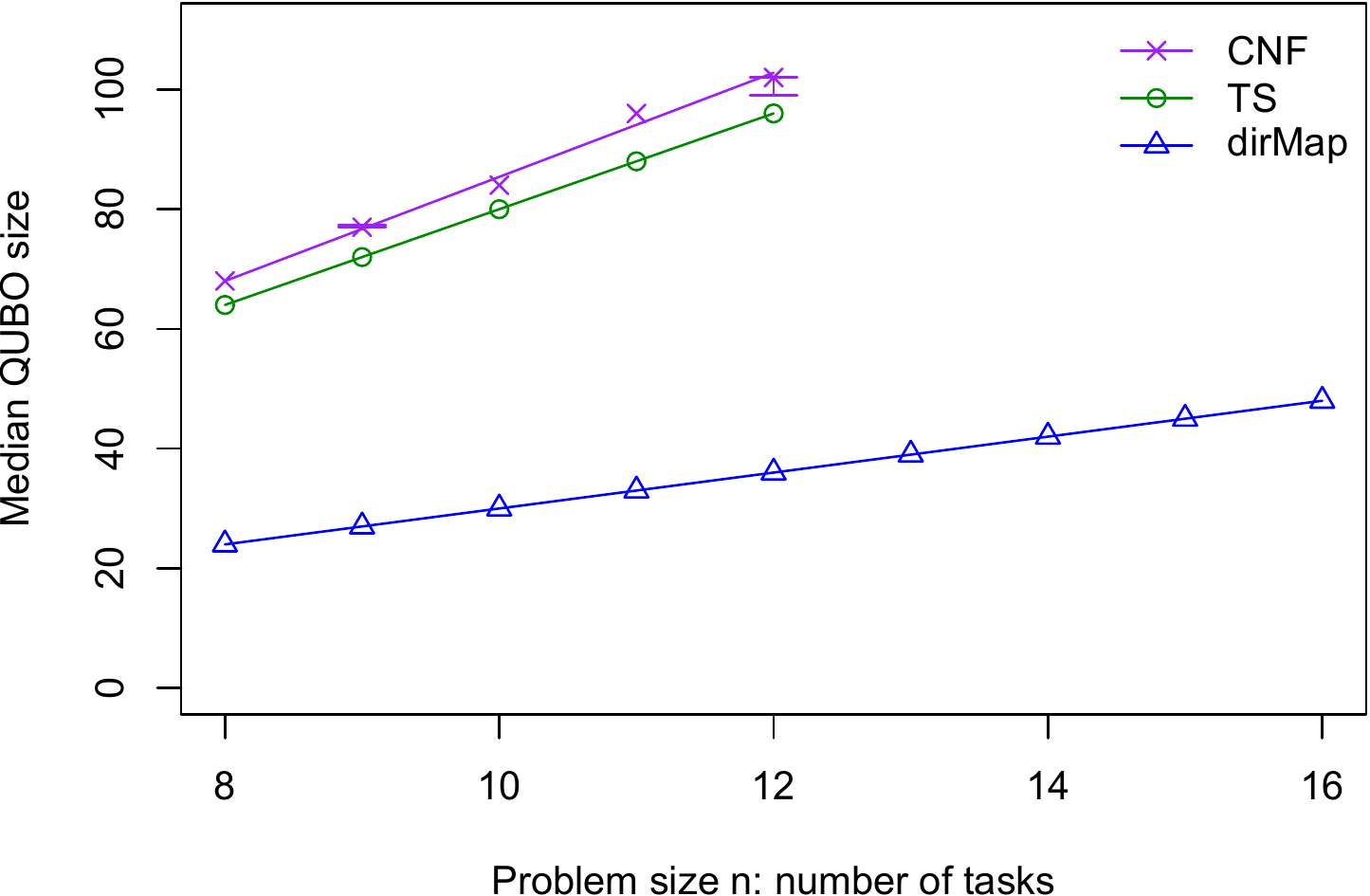}
\caption{{\bf Comparison of QUBO size across mappings of the scheduling-type 
planning problems.} The QUBO size
for the direct map is simply $3n$, and for the time-slice map it is $8n$,
where $n$ is the number of tasks that need to be scheduled. 
The QUBO sizes for the CNF mapping vary, so for this case, we are showing
the median QUBO size over the $100$ problems of that size. Some of the 
error bars
for the CNF mapping at the $35$th and $65$th percentiles are too small to see.}
\label{fig:qubo_size_comp}
\end{figure}

\begin{figure}
\includegraphics[width=\columnwidth]{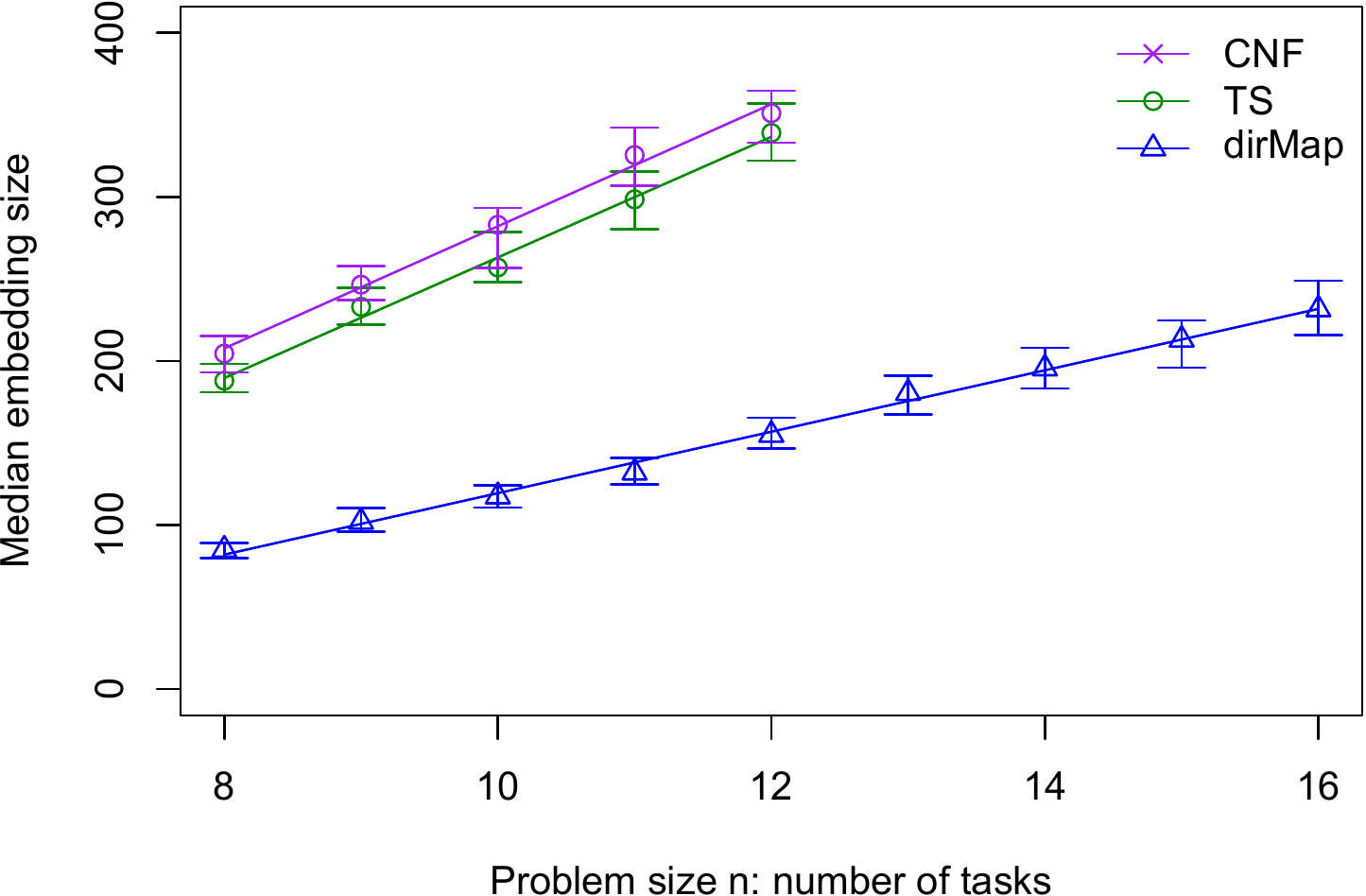}
\caption{{\bf Comparison of embedding size across mappings of the 
scheduling-type planning problems.} 
Median embedding size over the $100$ problems of size $n$. 
The error bars are at the $35$th and $65$th percentiles.}
\label{fig:embSizeVsProbSize}
\end{figure}

Not surprisingly, the median expected total anneal time to $99\%$ success 
is substantially lower
for the direct map instances than for the time-slice or CNF instances;
it is about a factor of $100$ smaller for all problem sizes in the range.
The direct mapping is tailored to 
this particular kind of scheduling-type planning problems,
rather than being applicable to planning problems in general. 
For this reason, the QUBO mapping is much more efficient 
(Fig.~\ref{fig:qubo_size_comp}), exactly $3/8$ smaller than the 
time-slice QUBOs across the entire size range. 
The typicial embedding size (Fig.~\ref{fig:embSizeVsProbSize}) is
also smaller, roughly half that of the other mappings. 
For these reasons, it is to be expected that the performance on 
the directly mapped problem instances is significantly better
than the performance for either of the general-purpose mappings
that, unlike the direct mapping, can be applied to any planning problem. 

\begin{figure}
\includegraphics[width=\columnwidth]{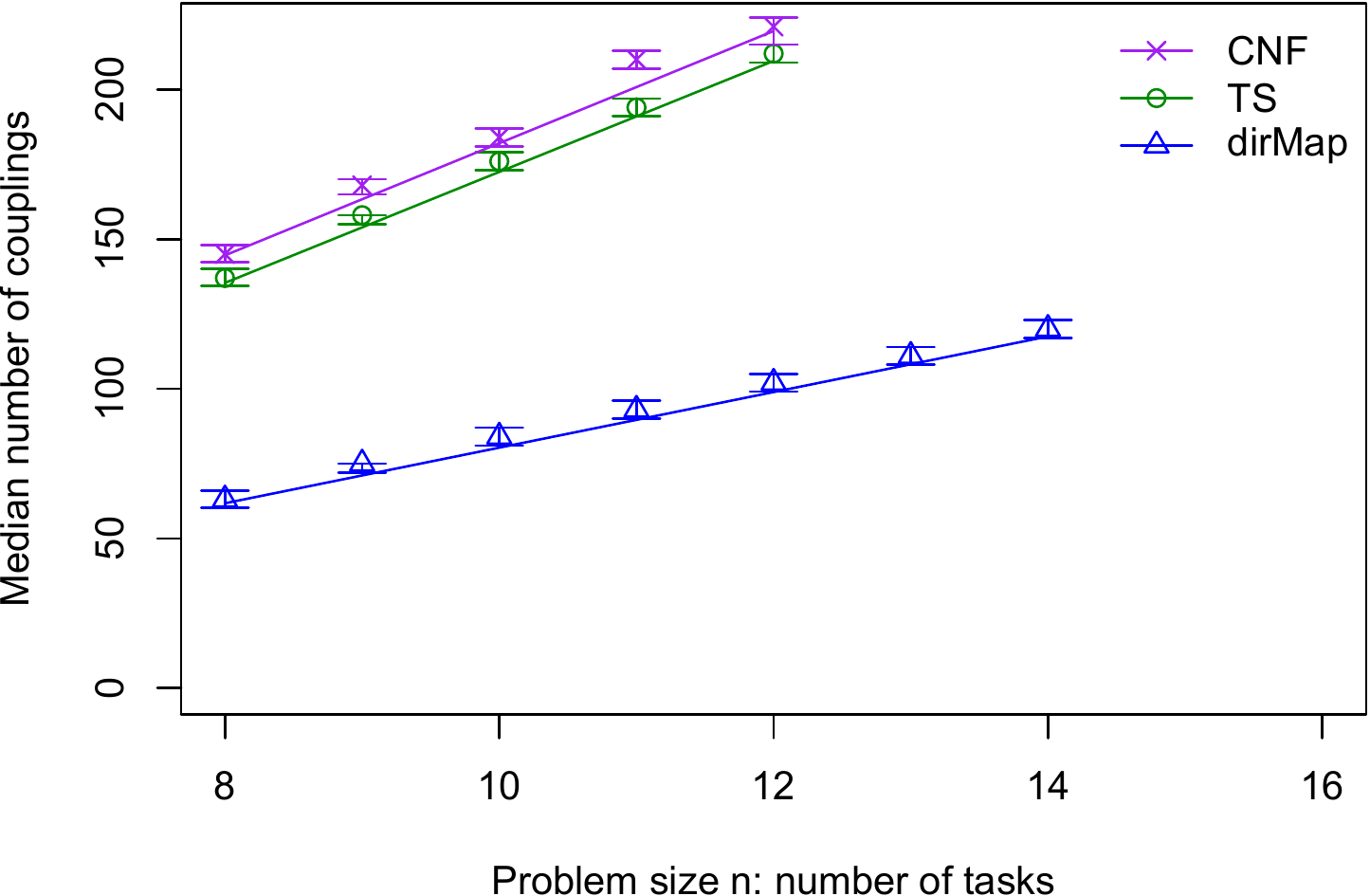}
\caption{{\bf Comparison of the number of couplings in the QUBOs obtained from the three different mappings of the scheduling-type planning problems.} 
The median number of couplings in the QUBOs for the $100$ problems of 
each size is shown, with error bars at the $35$th and $65$th percentiles.
}
\label{fig:couplingStats}
\end{figure}

More surprising is the substantial difference between the performance
on the time-slice mapped instances and the 
CNF mapped instances, with the median expected total annealing time 
to achieve $99\%$ success being about
a factor of $5$ greater for the CNF instances than the time-slice 
instances (Fig.~\ref{fig:bestComp}). The scaling for the time-slice
approach is also significantly better than for the CNF approach,
with an $\alpha$ value of $1.37$ rather than $1.76$
(though the scaling is estimated on very few data points). 
The time-slice and CNF mappings yield comparably-sized QUBOs
(Fig.~\ref{fig:qubo_size_comp}), with similar numbers of computings
(Fig.~\ref{fig:couplingStats}). For problem size $n$, the time-slice 
mapping yields a QUBO of size $8n$ qubits. 
The CNF approach yields variable size QUBOs, with the median size CNF QUBO
over $100$ problems only $4$ to $8$ qubits larger than the median size of
the time-slice QUBOs for problem sizes $\{8, 9, 10, 11, 12\}$.  
The median number of couplings for the CNF QUBOs exceeds that of the
time-slice QUBOs by only $\{8, 10, 8, 16, 9\}$  
for problem sizes $\{8, 9, 10, 11, 12\}$ respectively.  
Even the median embedding sizes of the CNF QUBOs are only
$\{7, 14, 26, 28, 12\}$ qubits larger respectively than the 
embedded time-slice QUBOs in this range, 
no more than a $10\%$ difference across this range
(Fig.~\ref{fig:embSizeVsProbSize}). 

\subsection{Comparing embedding properties across mappings}
\label{sec:embProps}

\begin{figure}
\includegraphics[width=\columnwidth]{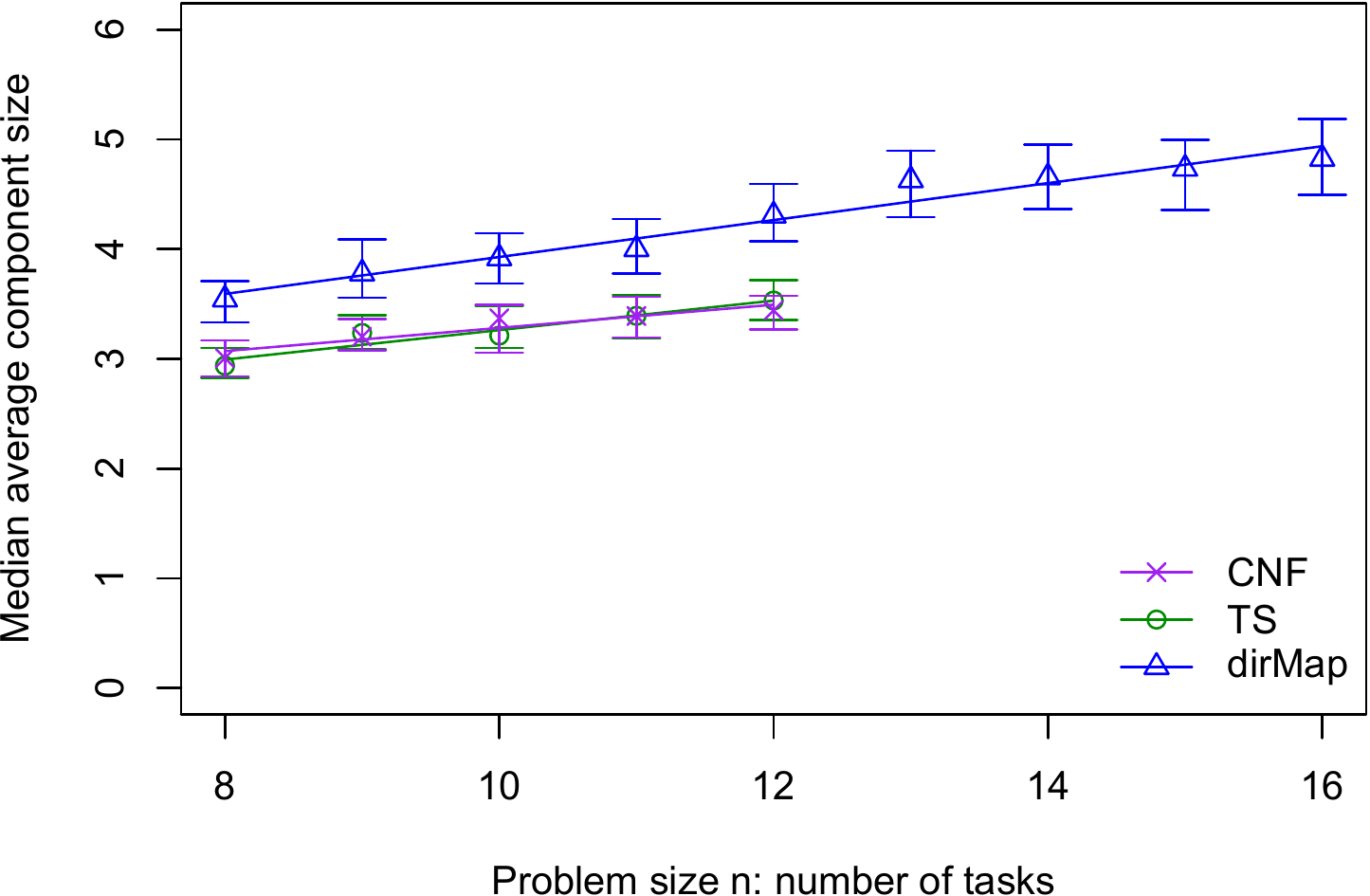}
% \caption{{\bf Comparison of average component sizes for 
% scheduling-type planning problems.} We computed the
% average of the component sizes, the number of physical qubits representing
% each logical qubit, and then took the median over all the $100$ problem
% instances of that size. Error bars are at the $35$th and $65$th percentiles.
% The median median component size (not shown) was $1$ for both the CNF and 
% time-slice mappings throughout the range considered. 
% The median median component size for the direct map, on the other hand,
% started at around $4$ at size $8$ and rose to close to $4.5$ by
% problem size $18$.
% }
% \label{fig:medAveCompSize}
% \end{figure}
% \begin{figure}
\includegraphics[width=\columnwidth]{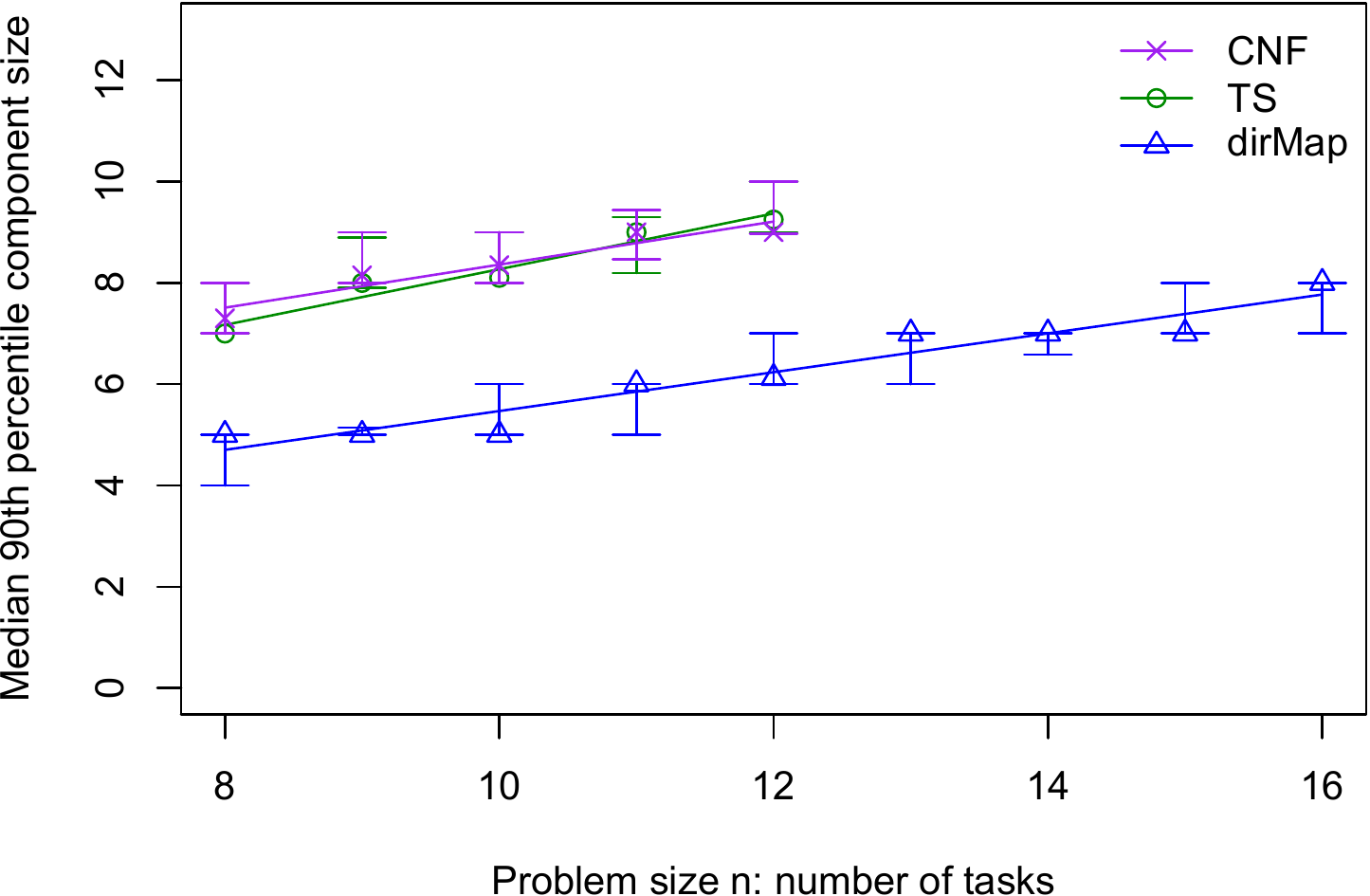}
% \begin{figure}
\includegraphics[width=\columnwidth]{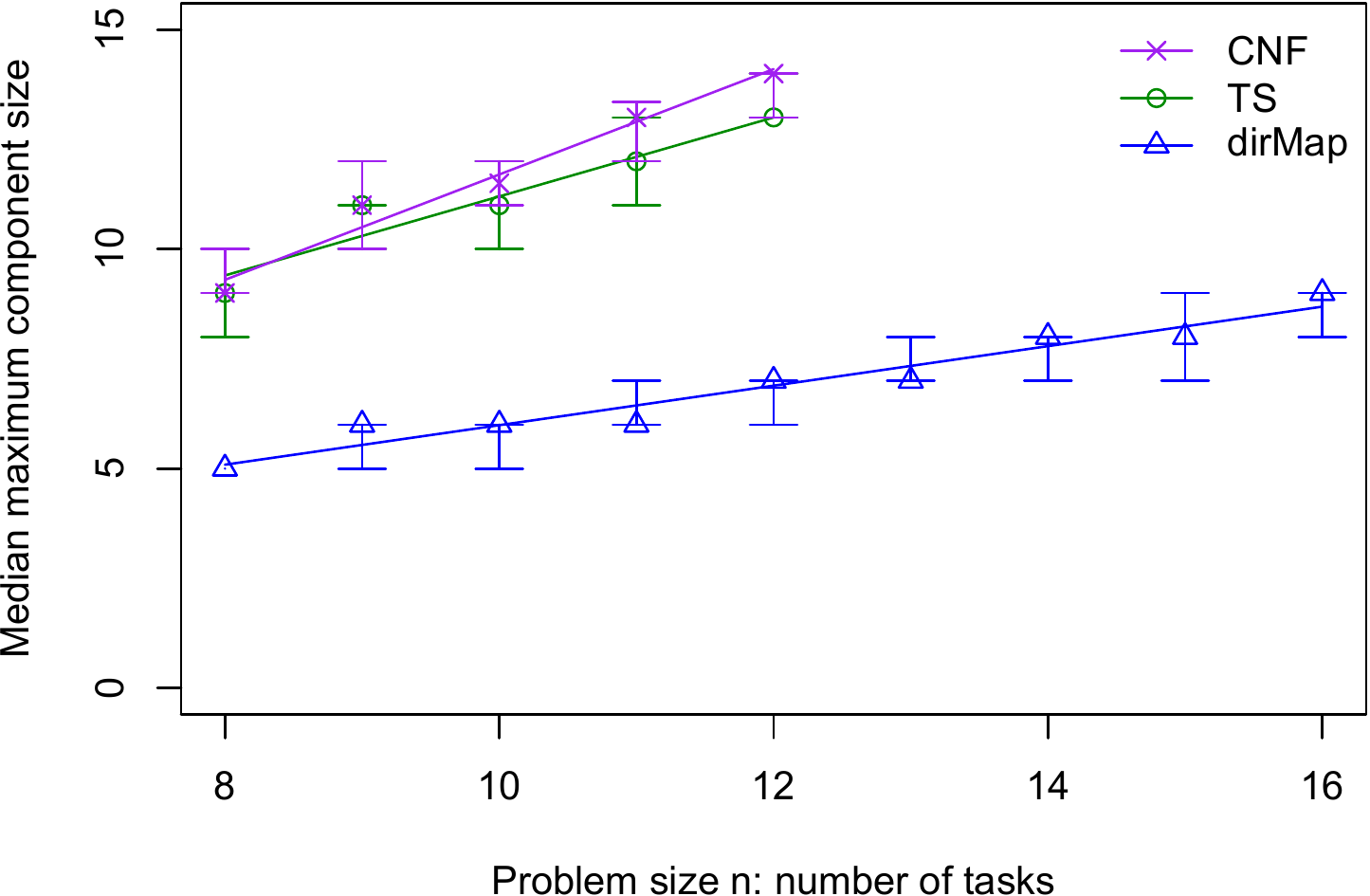}
% \caption{{\bf Comparison of maximum component sizes.} We determined the
% maximum component size for each problem, 
% and then took the median over all the $100$ problem
% instances of that size. Error bars are at the $35$th and $65$th percentiles.}
% \label{fig:maxCompVsProbSize}
% \end{figure}
\caption{{\bf Comparison of component sizes for scheduling-type planning
problems.} 
{\it Median average component size (Top)}: We computed the
average of the component sizes, the number of physical qubits representing
each logical qubit, and then took the median over all the $100$ problem
instances of that size.
{\it Median $90$th percentile component size (Middle)}: 
We calculated the component size at the $90$th percentile for each problem, 
and then took the median over all the $100$ problem
instances of that size. 
{\it Median maximum component size (Bottom)}:
We determined the largest component size for each problem, and then
took the median over all $100$ instances of each size.
Error bars are at the $35$th and $65$th percentiles.
}
% \label{fig:90CompVsProbSize}
\label{fig:componentComp}
\end{figure}

While the difference in performance between the CNF and time-slice 
instances is likely due in part to the somewhat larger size of the
embedded CNF QUBOs, other factors likely contribute to this difference. 
One possible factor is the size of the embedding components, the
number of physical qubits representing a logical qubit. For example,
D-Wave recommends both minimizing the size of the embedding
and the maximum component size in an embedding \cite{Cai-14}.
Also, for anti-ferromagnetic chains, Pudenz \ea~\cite{Pudenz2013error}
showed that the performance of the D-Wave Two decreases with chain size. 
Because in our case the components can have a more complex topology 
than a single chain, are subject to local fields, and have couplings 
to qubits outside the component, it is unclear to what extent we 
would see the same behavior. Thus, the relative component sizes require 
investigation. 

Fig.~\ref{fig:componentComp} (Top) confirms, as would be expected given
the similar enbedding and QUBO sizes for the CNF and time-slice
instances, that the 
median average component size across the $100$ problems 
is statistically indistiguishable in the two cases across the range of sizes.
Furthermore, throughout the size range tested, the median median component 
size -- the median over the $100$ problem instances of the median 
component size of each instance --
and even the median 65th percentile component size, for both mappings is $1$. 
By the 90th percentile, the component size has increased significantly
in both cases, but the two cases are statistically indistinguishable
(Fig.~\ref{fig:componentComp} (Middle)).
Even the median maximum component size hardly differs between the two
mappings (Fig.~\ref{fig:componentComp} (Bottom)): for problem sizes
$n = \{8, 9, 10, 11, 12\}$, the median maximum component size for 
the time-slice instances was $\{9, 11, 11, 12, 13\}$, whereas for the
CNF instances it was $\{9, 11, 11.5, 13, 14\}$. Given that the median 
maximum component size was identical for the two smallest sizes, even 
though the performance was markedly different in the two cases, and
the difference between the median maximum component sizes was no more than
$1$ for the other sizes, we conclude that component size did not
contribute significantly to the difference in performance between
the time-slice and CNF instances. 

Even though the QUBO size under the direct map is never much more than
roughly half the size of the QUBOs under the two general mappings
in the size range considered, both
in terms of number of qubits (Fig.~\ref{fig:qubo_size_comp}) and
the number of couplings (Fig.~\ref{fig:couplingStats}),
the typical component size in the embedded direct mapping instances
is markedly larger than for the CNF and time-slice mappings. 
Fig.~\ref{fig:componentComp} shows the median average component size. 
The direct map typical component size is significantly greater than the 
typical component size for the CNF and time-slice mappings throughout
the range of problem sizes.
Even at problem size $8$ it is markedly
higher than for the other two. 
While, the median median component size was $1$ for both the CNF and
time-slice mappings throughout the range considered,
the median median component size for the direct map
started at around $4$ at size $8$ and rose to close to $4.5$ by
problem size $18$.

The size of the top $10$ percent of the
components in these embeddings, however, is markedly lower than that
of the CNF and time-slice mappings; even at problem size $16$ it is 
still lower than the value for problem size $8$ for the more general mappings. 
These findings are consistent with D-Wave's recommendation to 
minimize the maximum component size, not just the total embedding
size (or equivalently, the typical component size). The results for
the time-slice versus CNF instances of the same underlying 
scheduling-type planning problems suggest that further investigation
is needed as to what properties of mappings and embeddings 
correspond to better or worse performance, and that more sophisticated
metrics for good embeddings are needed. 

\subsection{Comparing two different annealing profiles}

\begin{figure}
\includegraphics[width=\columnwidth]{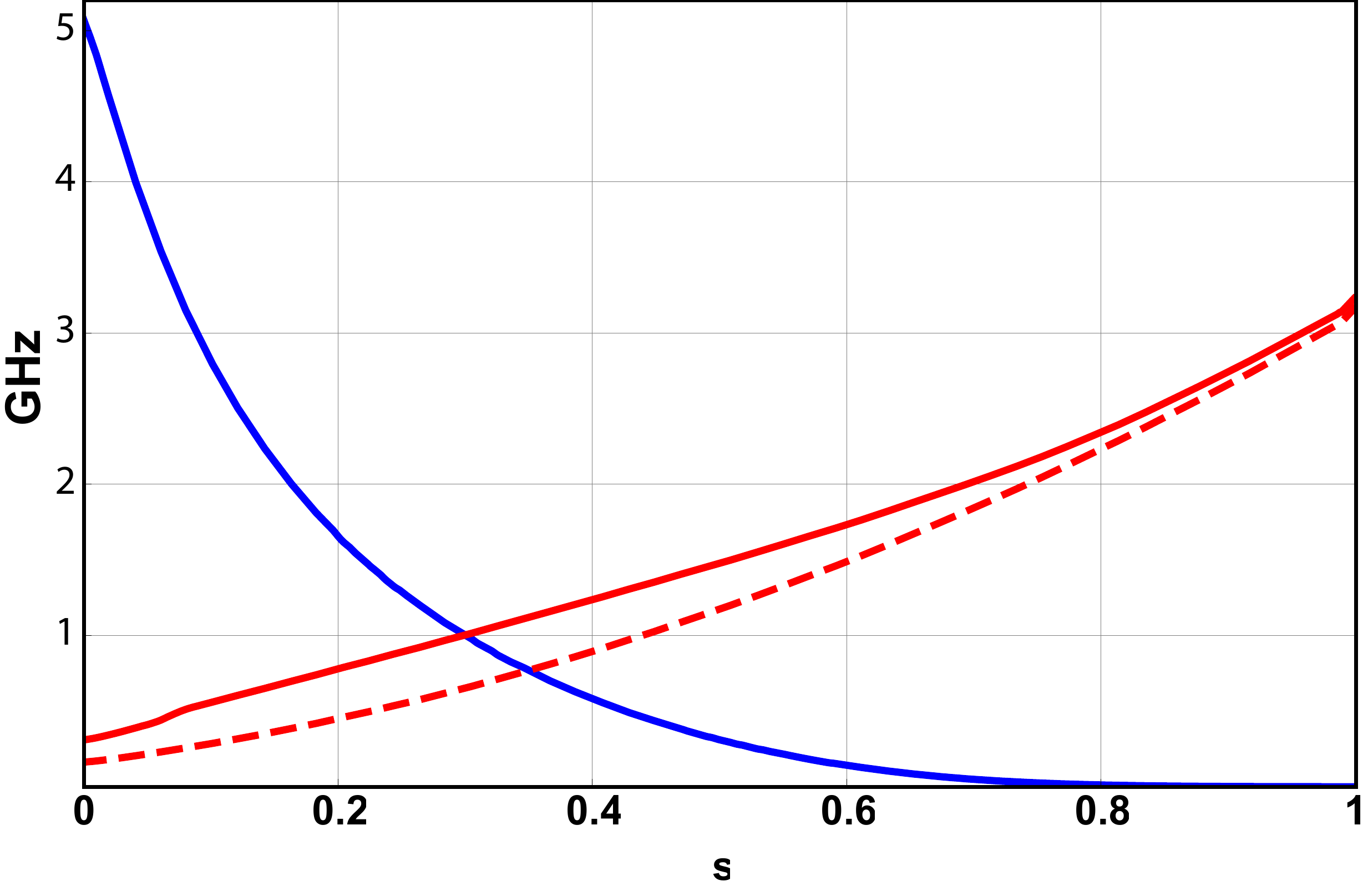}
\caption{{\bf Weightings of the different components of the Hamiltonian
prior to recalibration.} The figure shows the discrepancy
between the strength $B'(s)$ of the local fields (dashed line) and the 
strength $B(s)$ of the couplings (solid line) in the problem Hamiltonian 
during annealing prior to recalibration as they increase throughout
the anneal. The strength of the driver Hamiltonian as it decreases 
throughout the anneal is also shown. Normally in an anneal, there
is only a single weighting for the problem Hamiltonian,
as in Eqn~\ref{eqn:QAeqn}. Indeed, after recalibration, the strength 
of the local fields has been adjusted to equal that of the couplings,
so $B'(s)$ becomes $B(s)$.}
\label{fig:HamWeightings}
\end{figure}

The D-Wave Two machine has fixed functions $A(s)$ and $B(s)$ for the
weights of the driver and problem Hamiltonians in the
control Hamiltonian of Eqn.~\ref{eqn:QAeqn}.
The user can vary only the overall annealing time, and as mentioned,
the overwhelming evidence is that for the present machine the shortest
possible annealing time is best. Inadvertently, we were able to experiment
with an alternate annealing profile, albeit a nonstandard one. 
The D-Wave Two has two annealing lines, both common to all qubits. These 
lines affect the local fields $h_i$ and coupling coefficients $J_{ij}$
differently.
When they are not fully synchronized, effectively the $B(s)$ weighting
splits into two weightings, the original $B(s)$ weighting for the couplings
and a $B'(s)$ weighting for the local fields. When we first ran, 
these two lines were not fully synchronized. 
% with the effect that
% the weighting for the local fields was lower in strength than
% the weighting of the couplings, particularly in the first part of the
% anneal 
While the two converged near 
the end of the anneal, for much of the time, the
strength of the local fields for the problem Hamiltonian was $2-3$ GHz 
less than that of the coupling strength for the problem Hamiltonian 
(Fig.~\ref{fig:HamWeightings}). 
The synchronization issue was fixed by estimating the time-dependence 
of the persistent current and modifying the signal in annealing lines  
so as to compensate for the effect and make the weights on the 
coupling and local fields uniform.

\begin{figure}
\includegraphics[width=\columnwidth]{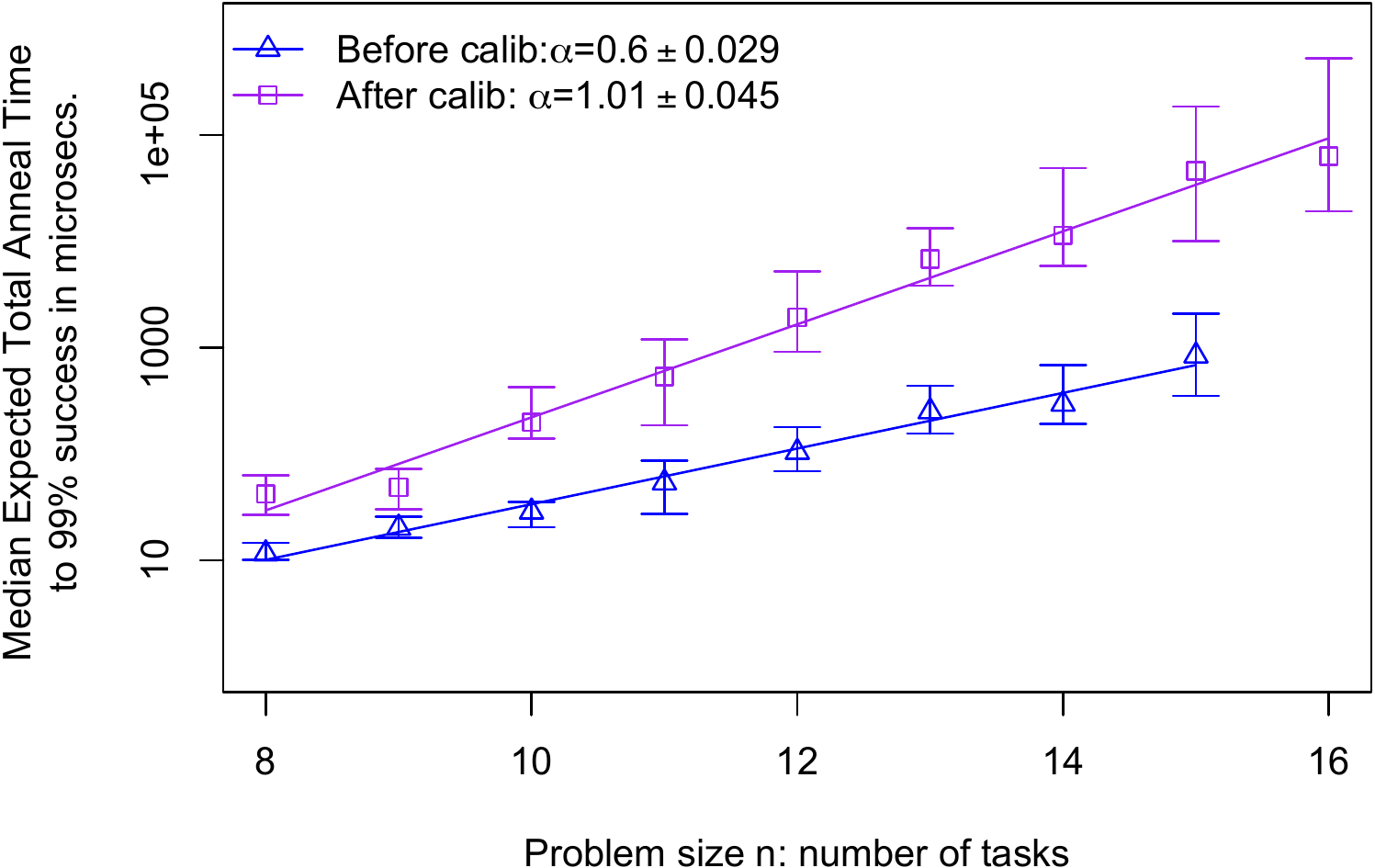}
\caption{{\bf Median expected total anneal time before and after a
recalibration.} 
The performance on the directly mapped scheduling-type planning problems 
decreased significantly after recalibration.
Shown is the median expected total anneal time
for $99\%$ percent success for $J_{int} = 1.25$ before the recalibration,
and the results for the best $J_{int}$ after calibration 
(same as in Fig.~\ref{fig:bestComp}).
% , both without error correction. 
The variability in the expected total anneal time across the $100$ problems
also increased, as illustrated by the error bars at the 35th and 65th 
percentiles.
While the recalibration improved results for other 
problems run on the machine, the reverse held for the planning problems
we ran.}
\label{fig:beforeAndAfterCalib}
\end{figure}

Fig.~\ref{fig:beforeAndAfterCalib} shows results on directly mapped
instances both before and after the recalibration. 
% we obtained while the machine was in this configuration are shown in 
While the recalibration improved
results for problems other groups were running on the machine, 
it resulted in a substantial decrease in
performance on the directly mapped scheduling-type planning problems, 
%\cite{}, 
both in terms of the absolute total
annealing time and in the scaling, with $\alpha$ increasing from $0.6$
to $1.0$. 

\subsection{Performance dependence on the internal coupling}
\label{sec:Jint}

\begin{figure*}
\includegraphics[width = 2.4\columnwidth]{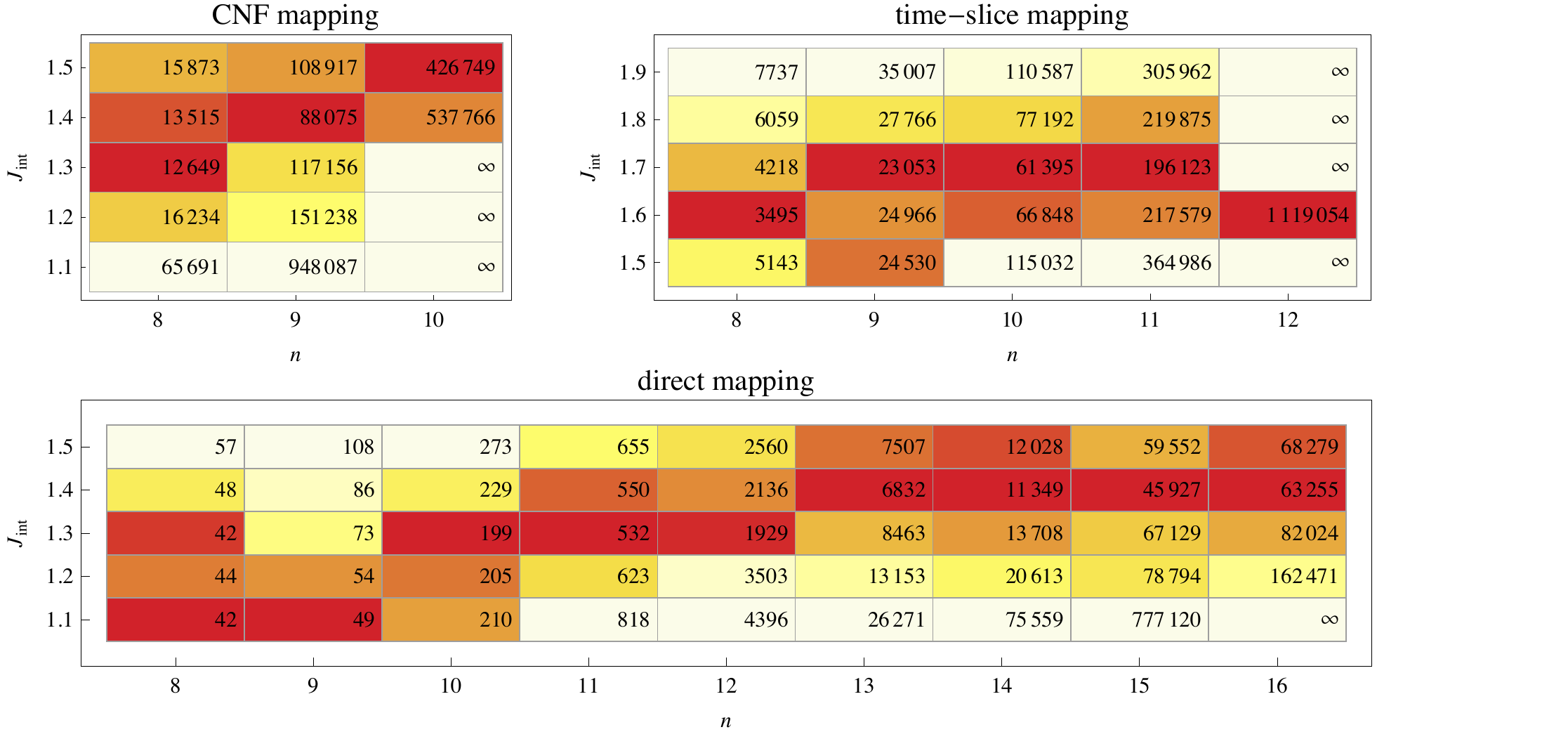}
\caption{{\bf Performance
for different values of $J_{int}$ on the three different mappings of 
scheduling-type planning problems.}
The numbers are the median expected total anneal time 
for $99\%$ percent success over the benchmark
set of $100$ hard but solvable scheduling-type problems of each size
for different values of $J_{int}$ under the three mappings.
% under the CNF mapping, time-slice mapping, and direct mapping. 
The colors were determined from
the success probabilities, with each column normalized by the largest
value for the given mapping and size. Deep red indicates the largest
value, with the colors ranging from orange to yellow to white indicating
lower and lower success probabilities.
}
\label{fig:Jint}
\end{figure*}

We now turn to how the value of $J_{int}$ affects performance for
each of the three different mappings.
Naively, one might expect that one should set $J_{int}$ as high as 
possible, so as to penalize states in which the bit values of the
physical qubits representing one logical qubit differ, but there are 
reasons why setting it too high can be detrimental \cite{Choi08}.
The first is that doing so can increase the difficulty of transfering
amplitude from a local minimum in which all of the bit values
for a component are equal but set to the wrong value.
The second is that the device has a finite range of possible couplings, and
suffers from significant precision issues, 
meaning that the noise in the applied coupling strengths is high enough
that only about $16$ different values can be distinguished. When an Ising 
problem is sent to the machine, the problem Hamiltonian is automatically 
rescaled so as to take advantage of the full range. 
If $J_{int}$ is set too high, however, when the problem 
is rescaled, all of the other field strengths become indistinguishable from 
$0$, and all information specific to the problem is lost. For these reasons,
there is a sweet spot for $J_{int}$. 

For each mapping, a quick parameter sweep was done (not shown) to 
locate the correct range in which to perform a more detailed evaluation.
The tables in Fig.~\ref{fig:Jint} give the median expected total anneal
time for $99\%$ percent success over the benchmark problems
under various values of $J_{int}$ for each of the three mappings,
with cell colorings determined from the success probabilities.
The table for the direct mapping shows that the optimal
value of $J_{int}$ increases as the problem size and the component size
increase, likely because a higher value of $J_{int}$ is useful for
pressuring all of the qubits in the larger components to take on the same
value by the end of the computation. The tables for the CNF and 
time-slice mapping are less conclusive, particularly since only
a handful of anneals resulted in a solution for the largest sizes (the last
column of each table), but they are consistent with this trend. The
time-slice instances benefit from somewhat higher values of $J_{int}$ than
the other two mappings. The same trend was also found in \cite{Venturelli14}.

\subsection{Performance with a simple error correction scheme}
\label{sec:EC}

\begin{figure}
\includegraphics[width=\columnwidth]{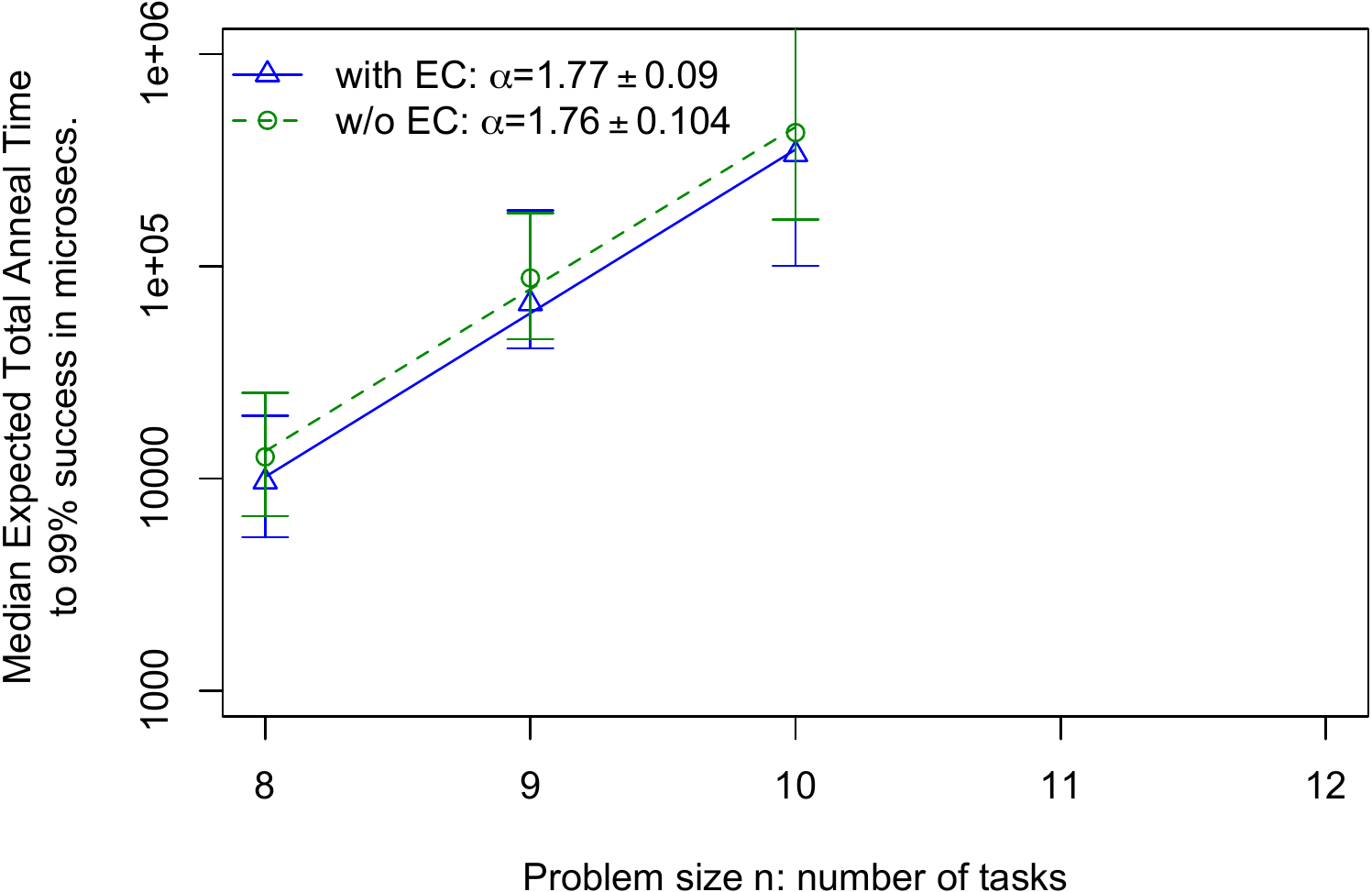}
\includegraphics[width=\columnwidth]{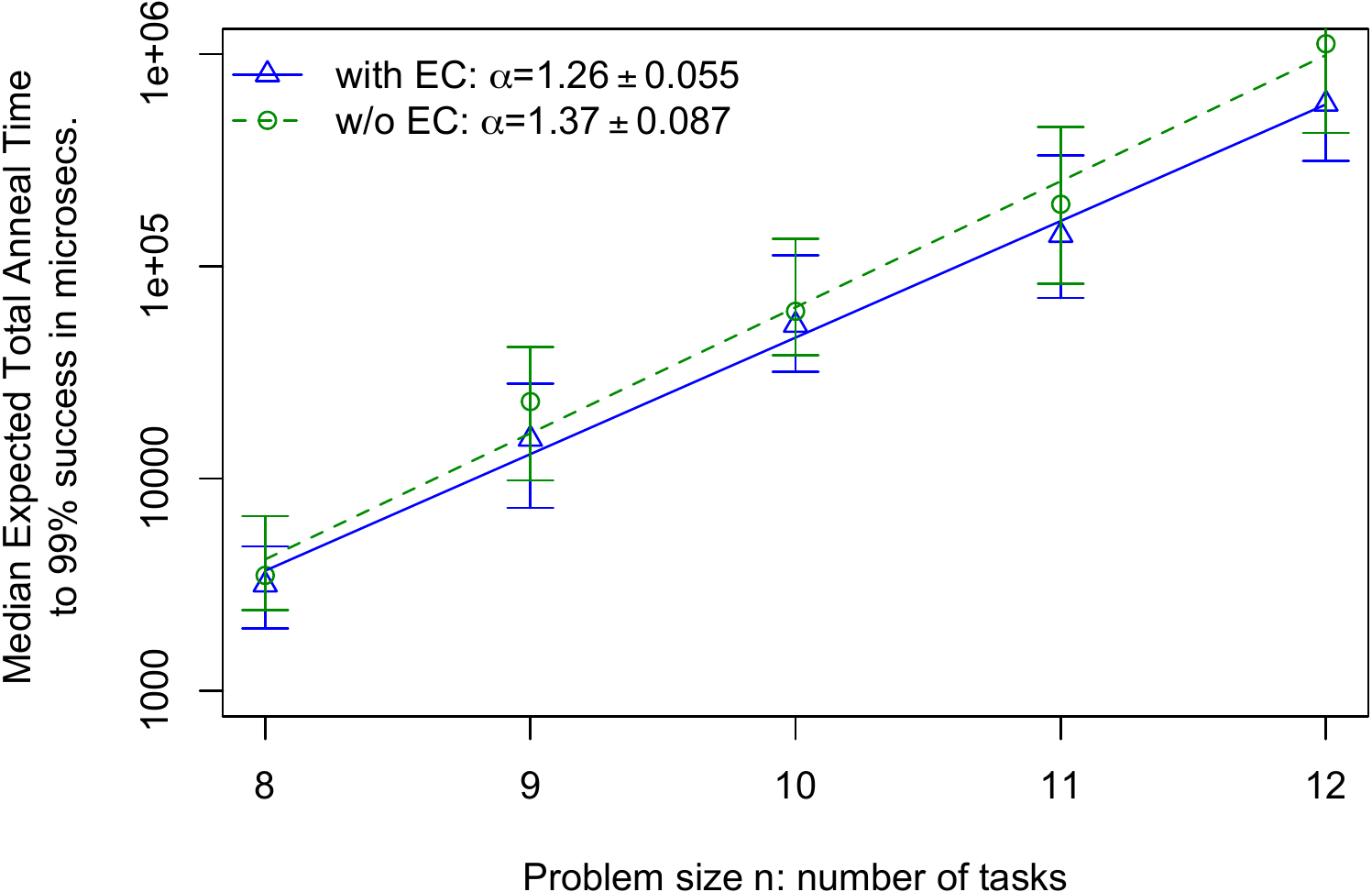}
\includegraphics[width=\columnwidth]{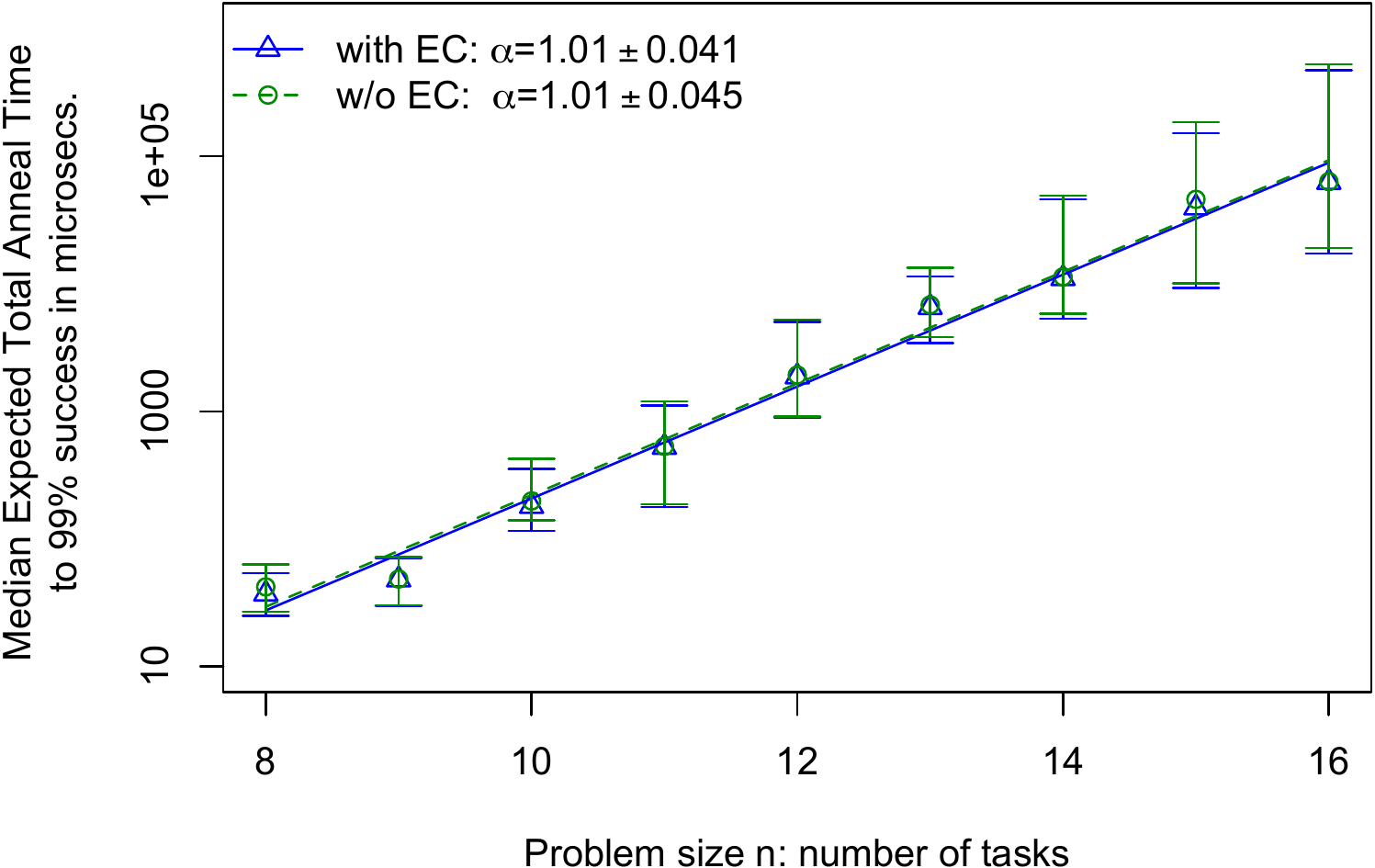}
\caption{{\bf Comparison of performance with and without simple 
error correction on the three mappings of scheduling-type planning problems.}
Median expected total anneal time for $99\%$ percent success 
for each mapping, with the best $J_{int}$ for each size, are shown. 
{\it CNF mapping (Top):}
In the corrected case, $J_{int}$ was $\{1.6, 1.5, 1.7, 1.5, 1.6\}$ for 
problem sizes $\{8, 9, 10, 11, 12 \}$ respectively. For the uncorrected case,
$J_{int}$ was $\{1.6, 1.7, 1.7, 1.7, 1.6\}$ respectively.
{\it Time-slice mapping (Middle):}
In the corrected case, $J_{int}$ was $\{1.2, 1.3, 1.3\}$ for 
problem sizes $\{8, 9, 10\}$ respectively. For the uncorrected case,
$J_{int}$ was $\{1.2, 1.3, 1.4\}$ respectively.
{\it Direct mapping (Bottom):}
In the corrected case, $J_{int}$ was 
$\{1.1, 1.1, 1.2, 1.3, 1.3, 1.4, 1.4, 1.4, 1.4\}$ for 
problem sizes $\{8, \dots, 12 \}$ respectively. For the uncorrected case,
$J_{int}$ was $\{1.1, 1.1, 1.3, 1.3, 1.3, 1.4, 1.4, 1.4, 1.4\}$ respectively.
}
\label{fig:EC}
\end{figure}

The simple error correction scheme of Sec.\ref{sec:methods}
does not provide significant improvements in performance on 
these problems under any of the mappings, as can be seen in
Fig.~\ref{fig:EC}. 
The time-slice results suggest that error correction could provide 
more significant improvements as the problem size increases.
The results on the directly mapped instances, however, show no
improvement with error correction over the whole range of problem sizes.
The error correction scheme acts only on physical qubits
representing the same logical bit that end up taking on opposite 
values. The reason for these opposite values is likely that there
are conflicting constraints with neighboring values. For this reason,
majority voting error correction at this level, while it removes any 
energy penalties 
resulting from disagreements within the component, will often 
increase the penalties from constraints involving neighboring
logical qubits.

% \begin{figure}
% \caption{{\bf Median expected total anneal time for $99\%$ success for the
% CNF mapping.} The results with the best $J_{int}$ are shown. In
% }
% \label{fig:bestCNF}
% \end{figure}
% \begin{figure}
% \caption{{\bf Median expected total anneal time for $99\%$ percent success 
% for the direct mapping.} 
% }
% \label{fig:bestDirMap}
% \end{figure}

\section{Results on navigation-type planning problems}
\label{sec:navResults}

For both the CNF and time-slice mappings of problems from the parametrized
family of navigation-type planning problems, the largest problem size that 
embedded was size $4$. For the navigation-specific direct mapping, 
problems of size $6$ embedded reliably, but problems of size $7$ no 
longer embedded. We increased the number of embedding tries, and even 
asked D-Wave to try to embed these problems for us, but without success. 
Only problems with fewer than $33$ vertices are guaranteed to embed
in the $512$ vertex $(8,4)$-Chimera graph, so these $49$-qubit QUBOs
may not embed in the current hardware. While the directly mapped
navigation-type planning problems have QUBO graphs that are far from
fully-connected - the number of edges in these graphs goes up linearly
with size not quadratically - they appear to be sufficiently connected
that it may be impossible to embed any size $7$ navigation-type planning
problem in the current hardware. The next section explores embedding
these problems in larger architectures.
For those problems that did embed, 
% \footnote{
% In addition to running problems from the navigation-type parametrized families,
% we also ran tiny problems inspired by data for 
% the Resource Prospector Mission (RPM) \cite{},
% a NASA mission currently being planned to explore permanently shadowed areas
% of the lunar polar region for ice. From a list of potential
% targets, and terrain maps of the region, we estimated whether the
% direct path between two targets is traversable. From the resulting graph,
% we obtained Hamiltonian path problems on subsets of the graph, and ran
% those. At the small sizes at which these embedded, the D-Wave machine had
% not trouble solving these problems.
% }
the D-Wave machine solved them reliably and quickly.

\section{Embedding in Future Architectures}
\label{sec:embResults}

\begin{figure}
\begin{center}
\includegraphics[width=1.0\columnwidth]{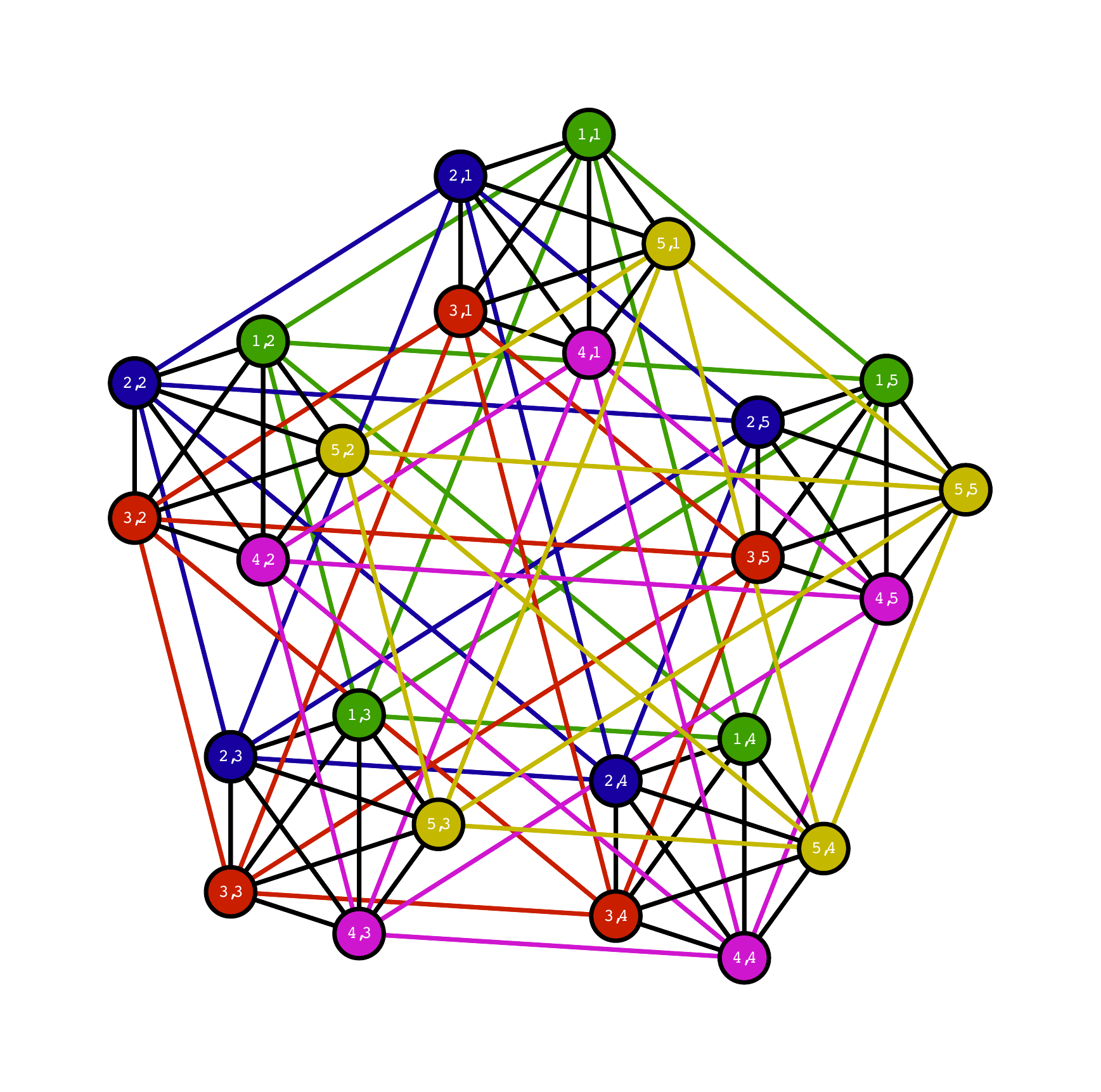}
\includegraphics[width=0.8\columnwidth]{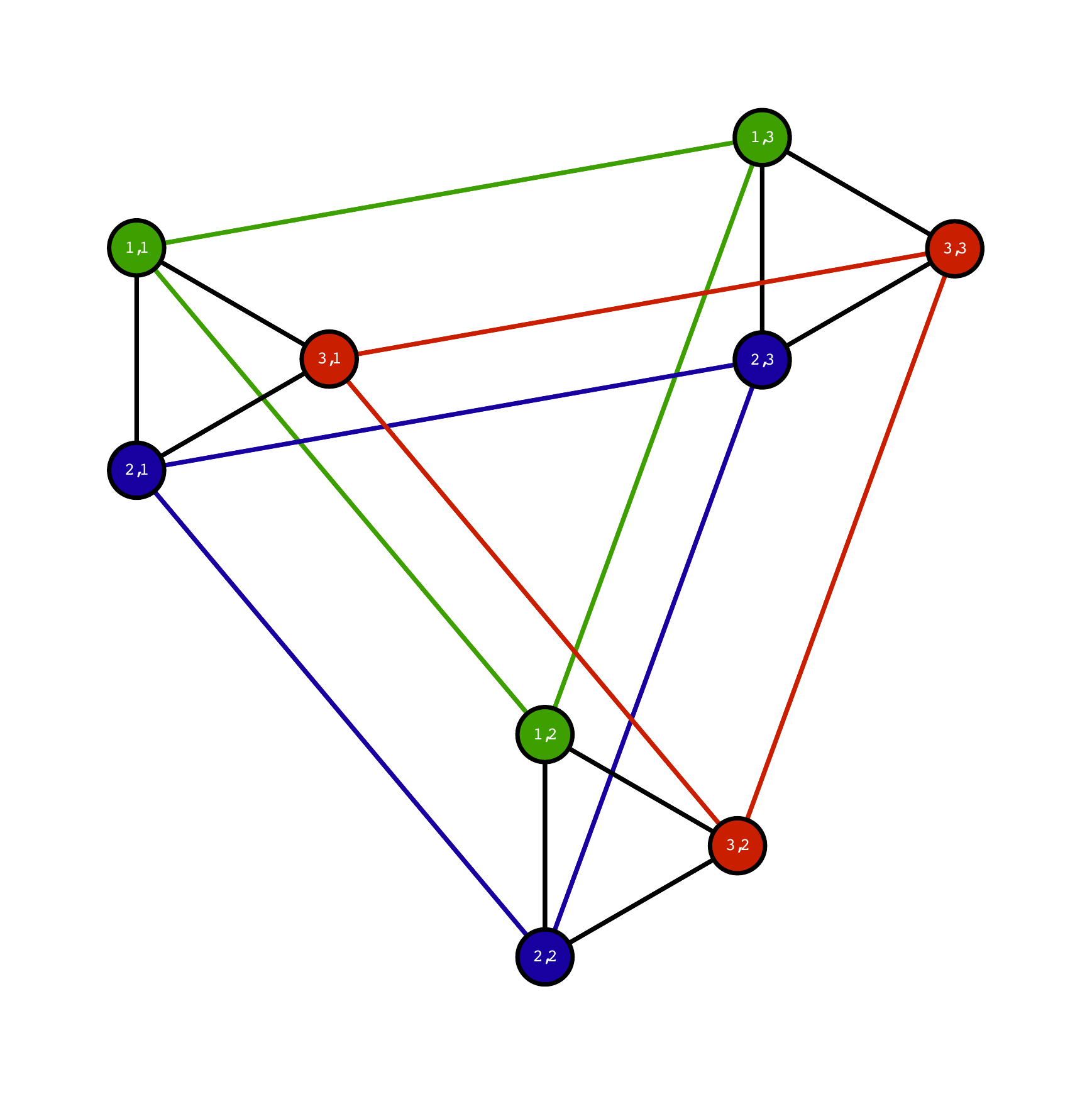}
\caption{{\bf The intersecting-cliques graphs $IC_{5,5}$ and $IC_{3,3}$.} The 
intersecting-cliques graph $IC_{n,n}$ is a subgraph of all navigation-type
problems with $n$ or more sites. It corresponds to the trivial navigation-type
problem in which every site is connected directly to every other site. It
is the most embeddable of the navigation-type problems, meaning that if
it does not embed then neither will any of the other problems with the
same number of sites (or more).} 
\label{fig:intCliques}
\end{center}
\end{figure}

We investigated the embeddability
of the most embeddable graph, the graph corresponding to the direct mapping
to QUBO of the trivial Hamiltonian path problem on the fully connected graph
$K_n$ of size $n$. This graph, which we call the 
intersecting-cliques graph $IC_{n,n}$
for reasons that will become apparent shortly, is a subgraph of all
the directly mapped navigation problems of its size (and larger) so if this
one doesn't embed, none of the others will. The graph is far from fully 
connected, each of the $n^2$ vertices having degree only $2(n-1)$,
but the degree does grow in linearly with the size of the problem. 
The $n^2$ vertices in the graph can be partitioned into a set of $n$ 
cliques each containing $n$ vertices in two different ways. 
The first clique partition corresponds to the QUBO term enforcing the 
condition that each site is visited exactly once, and the second to 
the QUBO term enforcing the condition that only one site is visited 
at a time. The $n$ vertices in any one of the cliques in the first set 
are all in different cliques of the second set. This property inspired 
the name. 
Diagram of $IC_{5,5}$ and $IC_{3,3}$ are shown in Fig.~\ref{fig:intCliques}.

\begin{figure}
\begin{center}
\includegraphics[width=0.9\columnwidth]{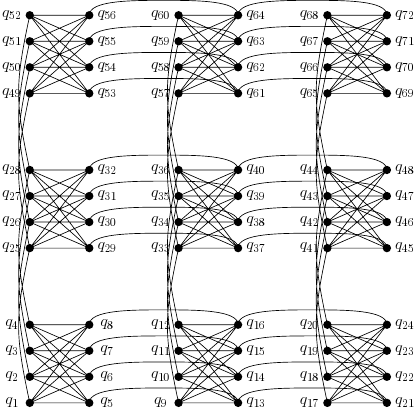}
\caption{{\bf The $(3,4)$-Chimera graph.} A schematic diagram 
from \cite{Smelyanskiy12} of
the $(M,L)$-Chimera graph underlying D-Wave's architecture. In the
$(3,4)$-Chimera graph shown, there are $M^2 = 9$ unit cells, each of 
which is a fully-connected bipartite graph $K_{4,4}$ containing $2L$
qubits. The qubits in the left column of each unit cell are
connected to the analogous qubits in the unit cells above and below
(add or subtract $ML = 24$ to the index) and 
the qubits in the right column of each unit cell are
connected to the analogous qubits in the unit cells to the right
and left (add or subtract $2L = 8$ to the index).
The D-Wave Two used in the experiments has a $(8,4)$-Chimera graph 
architecture, but with $3$ broken qubits that are not used. } 
\label{fig:ChimeraArch}
\end{center}
\end{figure}

We investigated the embeddability of these problems in potential
future architectures, specifically $(M,L)$-Chimera graphs 
\ref{fig:ChimeraArch}
in which either the number of unit cells $M^2$ or the size of the unit 
cell $K_{L,L}$ is increased, or both. 
A deterministic algorithm \cite{Klymko14} provides an embedding of 
any graph with no more than $ML + 1$ vertices in an $(M,L)$-Chimera graph, 
but for graphs that are far from fully connected, this algorithm is usually
quite inefficient in the number of qubits used, and in practice many
graphs can embed in a smaller Chimera graph than is found by this
algorithm. In particular, D-Wave's heuristic embedding software often
finds more efficient embeddings. 

To investigate the embeddability
of these problems in future architectures, 
we ran D-Wave's heuristic embedding software $11$ times per problem 
on each $(M,L)$-Chimera graph architecture. The heuristic embedding
software was run with default parameters, including up to $10$ restarts
if an embedding is not found on a given try. We recorded the $11$ 
embedding sizes we obtained. 

\begin{figure}
\begin{center}
\includegraphics[width=\columnwidth]{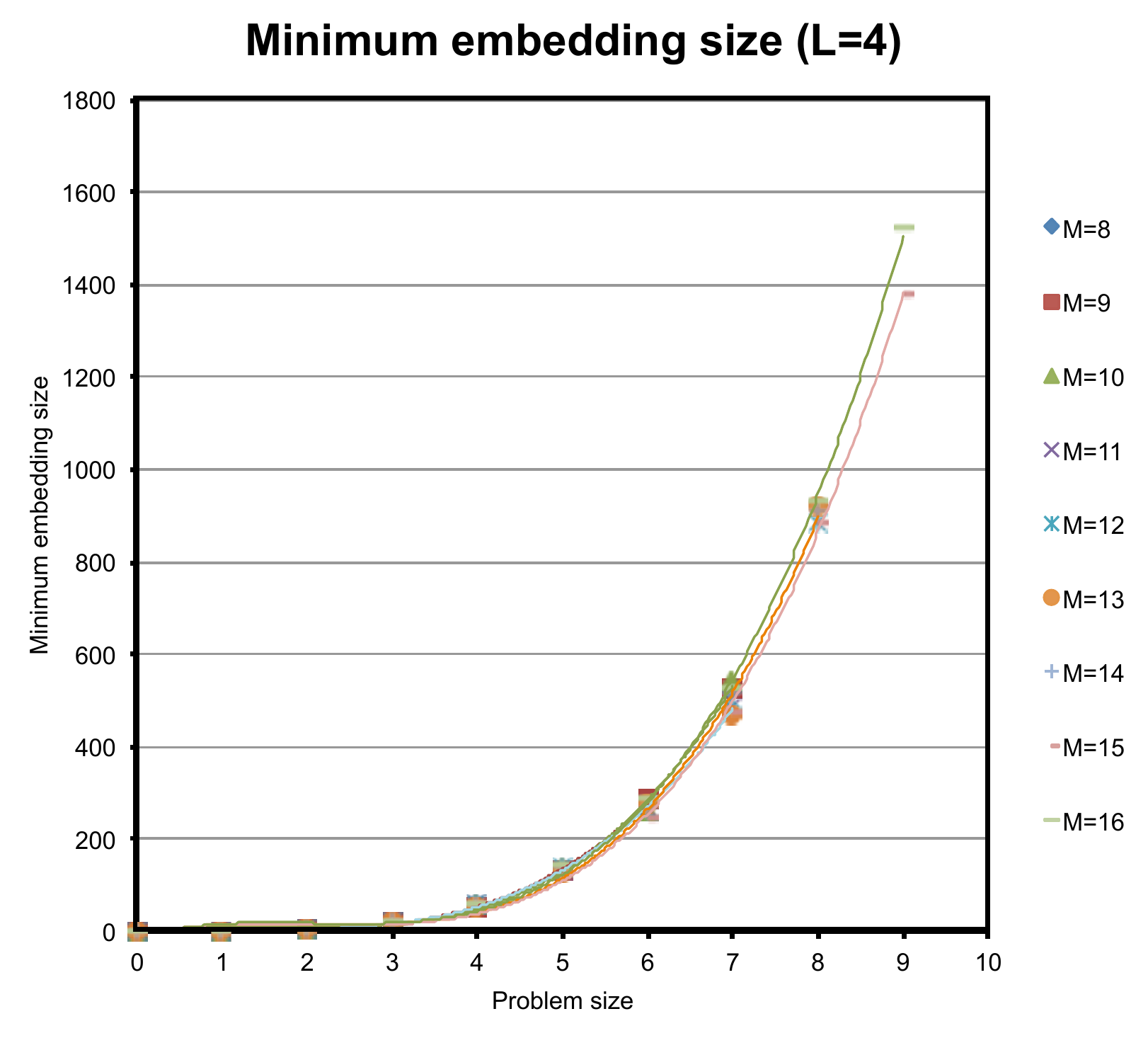}
\end{center}
\caption{{\bf $IC_{k,k}$ embedding sizes with increasing number of 
$K_{4,4}$ unit cells.} The embedding sizes of the intersecting-cliques graph 
$IC_{k,k}$ for $k \leq 10 $ in $(M,4)$-Chimera graphs for 
$M\in\{8, \dots, \}$. Increasing
the number of unit cells, $M^2$, does little to improve the embedding
sizes of graphs that embedded in a smaller architecture, but does
extend somewhat the range of $k$ over which $IC_{k,k}$ embeds 
reliably within $11$ runs of the D-Wave heuristic embedding software 
with default parameters. The points for sizes $0$ and $1$ were put
in manually. The rest reflect runs of D-Wave's heuristic embedding
software.}
\label{fig:L4graph}
\end{figure}

Fig.~\ref{fig:L4graph} shows the minimum embedding size for
architectures in which the size of the unit cell $K_{L,L}$ remains constant
with $L$ remaining 
at its current value of $L=4$, but the number of unit cells 
is increased. As can be seen in the figure,
increasing the number of unit cells,
while retaining the current size of the unit cells ($L=4$), hardly
improves the size of the embeddings of problems that embeddeded in
smaller architectures, but does extend the problem size that embeds
somewhat. 

\begin{figure}
\includegraphics[width=\columnwidth]{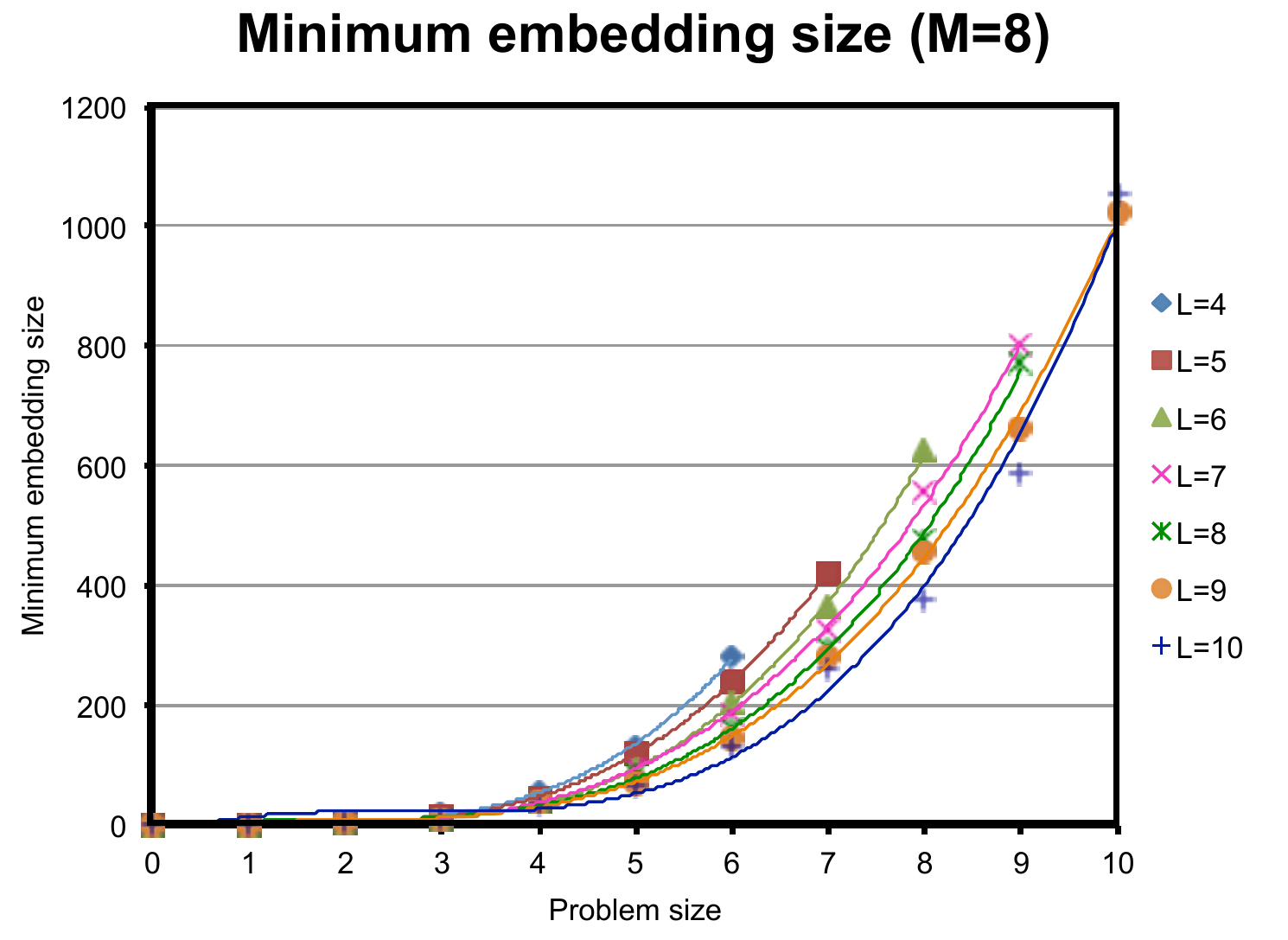}
\caption{{\bf $IC_{k,k}$ embedding sizes with increasing unit cell
size.} The embedding sizes of the intersecting-cliques graph 
$IC_{k,k}$ for $k \leq 10 $ in $(8,L)$-Chimera graphs for 
$L\in\{4,\dots,10\}$. Increasing
the size of the $K_{L,L}$ unit cells, and thereby the local connectivity of
the hardware graph,  while keeping the number
of unit cells constant, enables improved embedding 
sizes of graphs that embedded in a smaller architecture, as well as
extending the range of $k$ for which $IC_{k,k}$ embeds reliably.}
\label{fig:M8graph}
\end{figure}

Fig.~\ref{fig:M8graph} shows the minimum embedding size for
architectures in which the size of the unit cell $K_{L,L}$ is increased
while the number of unit cells stays constant at its
current value $M^2 = 64$. As can be seen in the figure,
increasing the size of the unit cells, and thus increasing the
local connectivity of the graph, even while the number of 
unit cells constant,
significantly improves the embedding size and the
embeddability of these problems. 

\begin{figure}
\includegraphics[width=\columnwidth]{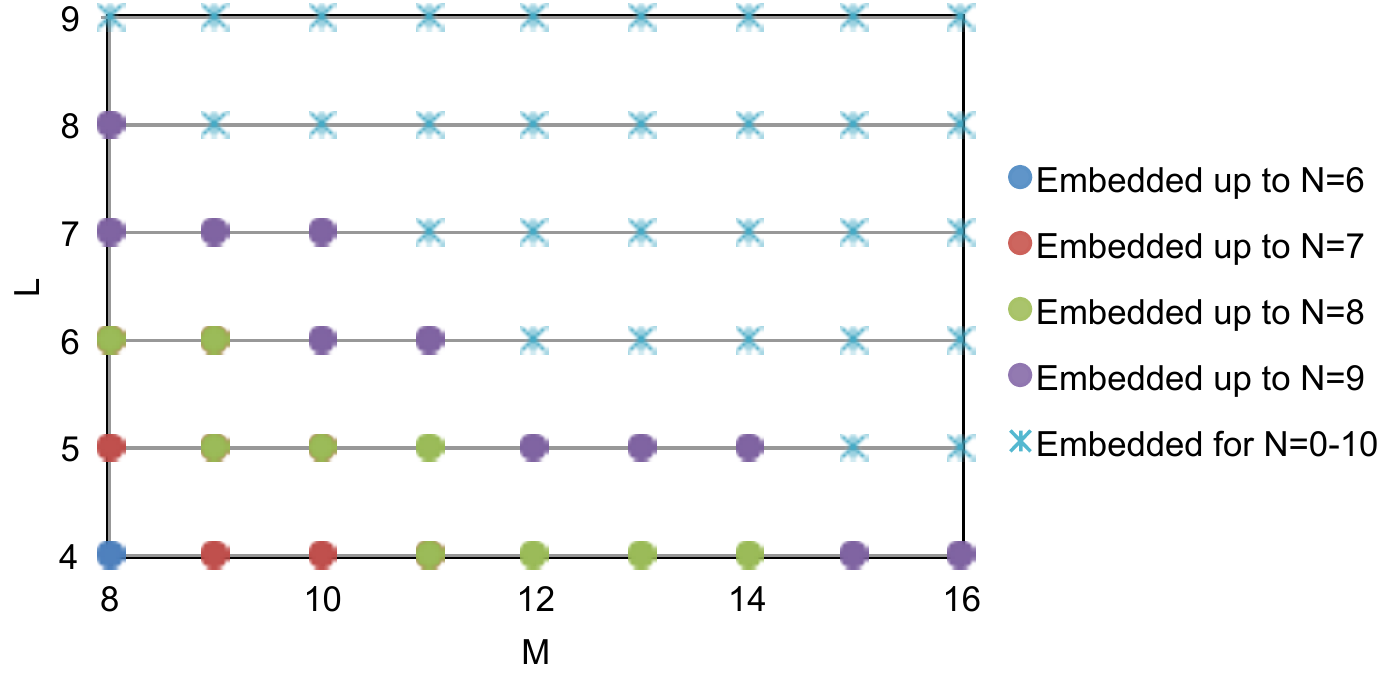}
\caption{{\bf Embedding of $IC_{k,k}$ graphs in $(M,L)$-Chimera
graphs.} Embedding of $IC_{k,k}$ for $k \leq 10$ 
in $(M,L)$-Chimera graphs for $M\in \{8,\dots,16\}$ and
$L\in\{4,\dots,8\}$. Dots show the largest size problem which embedded
in the given architecture in $11$ runs of the D-Wave heuristic embedding
software with default parameters. Crosses indicate
that the next largest size problem ($n = 11$) was not run, so we
do not have data to indicate whether or not it would have embedded
in the given Chimera architecture.}
\label{fig:MLcombo}
\end{figure}

Fig.~\ref{fig:MLcombo} explores the embeddability of $IC_{k,k}$ for
$k\in\{7,8,9,10\}$ for the full range of architectures with
$M\in\{8, \dots, 16\}$ and $L\in\{4, \dots, 10\}$.
The embedding software runs in seconds for problem sizes less than $7$,
but increases rapidly after that, with the $11$ embedding trials
taking $30-45$ minutes for some of the larger problem sizes. For this reason,
we did not explore problem sizes larger than $10$.

By the end of 2014, a $1024$ qubit D-Wave machine should be available
at NASA, and sometime in 2015, a $2048$ machine should be available. 
Unfortunately, if these qubit numbers are achieved by simply increasing
the number of unit cells, rather than increasing the connectivity of the
unit cells, or moving to a different architecture altogether, it looks 
likely that only navigation-type planning problems of size less than $10$
will be able to be run on such machines due to the difficulty embedding
larger instances in these hardware graphs.
% Fig.~\ref{fig:L4graph} shows that the intersecting-cliques graph $IC_{9,9}$
% did not embed in the $2048$ qubit $(16,4)$-Chimera. 
% Increasing the connectivity of the unit cells would provide greater, but 
% still limited, improvement. For example, 
% Fig.~\ref{fig:M8graph} shows that the intersecting-cliques graph $IC_{9,9}$
% embeds in the $1024$ qubit $(8,8)$-Chimera. 
Alternative embedding strategies could improve these results, but 
in the near-term, scheduling-type planning problems appear more 
suitable than navigation-type planning problems for these near-term
Chimera architectures with $K_{4,4}$ unit cell.
For real world planning applications that combine both navigation and 
scheduling aspects, the design of new hardware architectures
that will overcome the programming bottleneck presented by embedding
these problems is critical.

While the scheduling-type planning problems embed better than 
the navigation-type planning problems,
for both of the general mappings, by around problem size $15$ or $16$, 
the heuristic embedding software has difficulty embedding these problems. 
For the direct map instances, the heuristic embedding software had no
problems embedding instances up to problem size $17$, begins to have occasional
difficulties at problem size $18$, and has serious difficulties embedding
most of the problems by size $22$.

We finish the discussion of embedding in future architectures 
with an analysis of the embeddability of the simplest solvable 
scheduling-type planning 
problem, the one that corresponds to a completely disconnected graph of
$n$ vertices. As in the navigation case, its QUBO graph is a subgraph
of the QUBO graph for all of the scheduling-type problems, so if it
doesn't embed, none will. The analysis is easier than in the navigation 
case, and can be done by hand. 
The QUBO graph consists of $n$ triangles (we are considering the
$3$ time slot case). Since $k$ triangles
can embed in a $K_{2k,2k}$ unit cell, an $(M,2k)$-Chimera architecture 
% with an $M$ by $M$ grid of $K_{2k,2k}$ cells 
supports the embedding of the 
simplest $kM^2$-task problem. Since $4$ vertices are required 
to embed a triangle in Chimera graph architecture, this embedding
is optimal. As an example, the simplest problem with $128$ tasks
embeds in a $512$ qubit machine, and the $512$ task problem embeds
in the $2048$ machine projected for 2015. In addition, preliminary
tests show that typical scheduling-type problems of size well past $50$
will embed in the $2048$ architecture.

\section{Conclusions and Future Work}
\label{sec:concl}

We have studied the effectiveness of a quantum annealer in solving
small instances within families of hard operational planning problems
under various mappings and embeddings. From these
results we derive insights useful for the programming and design of
future quantum annealers: problem choice, the mapping used, properties
of the embedding, and annealing profile all matter, each significantly
affecting the performance. While this initial study did not produce
results competitive with state-of-the-art classical approaches, 
higher quality qubits, shorter annealing times, better precision, 
greater hardware connectivity, improved mappings and embeddings, and
alternative annealing profiles will all contribute to improved results.

In future work, we will experiment with other mappings, including 
experimenting with different weightings of penalty terms in the QUBO
funtion and other translations of the problem to CNF form, and also
other problems from operational planning, such as job shop scheduling. 
We will also build on this work, exploring more embeddings per problem 
to sort out how much of the variation
in runtime is due to the variation in the difficulty of the problems
themselves versus how much is due to properties of the embedding independent
of the problem. We will perform  statistical analysis of the correlation 
between a richer set of embedding properties and expected total annealing time
in order to give insight into better metrics for evaluating embeddings
of QUBO problems for quantum annealing. 

% \egr{Add somewhere need for general purpose planning algs, not just
% specific ones.}

% Scheduling problems preferable to navigation problems.
% 
% More uniform component sizes may make for better annealing
% solution times: the size of the top $10\%$ of the components may make
% a better figure of merit than the typical component size, and reducing
% that size at the expense of the typical component size may be a 
% profitable trade0ff.
% 
% Mapping matters.

\section*{Acknowledgements}
The authors are grateful to Jeremy Frank, Alejandro Perdomo-Ortiz, Ross Beyer,
Sergey Knysh, Chris Henze, and Itay Hen for helpful discussions,
and to DWave for technical support
and for discussions related to the calibration issue and our results before
and after.

\bibliographystyle{apalike} 
\bibliography{master,phaseTrans,qc}

% \appendix
% \section{Appendix}

% \input{./sections/appendix.tex}

\end{document}